\documentclass[12pt,preprint]{aastex}
\usepackage{multirow}
\usepackage{url}

\newcommand{\be}{\begin{equation}}
\newcommand{\ee}{\end{equation}}
\newcommand{\ba}{\begin{eqnarray}}
\newcommand{\ea}{\end{eqnarray}}

\def\g{\gamma}
\def\G{\Gamma}

\slugcomment{{\it The Astrophysical Journal}, accepted version}

\shorttitle{{\it Fermi}-LAT Constraints on the Gamma-ray Opacity of the Universe}
\shortauthors{The Fermi LAT \& GBM Collaboration}

\title{{\em Fermi} Large Area Telescope Constraints on the Gamma-ray Opacity of the Universe}

\author{
A.~A.~Abdo\altaffilmark{1,2}, 
M.~Ackermann\altaffilmark{3}, 
M.~Ajello\altaffilmark{3}, 
A.~Allafort\altaffilmark{3}, 
W.~B.~Atwood\altaffilmark{4}, 
L.~Baldini\altaffilmark{5}, 
J.~Ballet\altaffilmark{6}, 
G.~Barbiellini\altaffilmark{7,8}, 
M.~G.~Baring\altaffilmark{9}, 
D.~Bastieri\altaffilmark{10,11}
% ...
}

% ...
\altaffiltext{1}{Space Science Division, Naval Research Laboratory, Washington, DC 20375, USA}
\altaffiltext{2}{National Research Council Research Associate, National Academy of Sciences, Washington, DC 20001, USA}
\altaffiltext{3}{W. W. Hansen Experimental Physics Laboratory, Kavli Institute for Particle Astrophysics and Cosmology, Department of Physics and SLAC National Accelerator Laboratory, Stanford University, Stanford, CA 94305, USA}
\altaffiltext{4}{Santa Cruz Institute for Particle Physics, Department of Physics and Department of Astronomy and Astrophysics, University of California at Santa Cruz, Santa Cruz, CA 95064, USA}
\altaffiltext{5}{Istituto Nazionale di Fisica Nucleare, Sezione di Pisa, I-56127 Pisa, Italy}
\altaffiltext{6}{Laboratoire AIM, CEA-IRFU/CNRS/Universit\'e Paris Diderot, Service d'Astrophysique, CEA Saclay, 91191 Gif sur Yvette, France}
\altaffiltext{7}{Istituto Nazionale di Fisica Nucleare, Sezione di Trieste, I-34127 Trieste, Italy}
\altaffiltext{8}{Dipartimento di Fisica, Universit\`a di Trieste, I-34127 Trieste, Italy}
\altaffiltext{9}{Rice University, Department of Physics and Astronomy, MS-108, P. O. Box 1892, Houston, TX 77251, USA}
\altaffiltext{10}{Istituto Nazionale di Fisica Nucleare, Sezione di Padova, I-35131 Padova, Italy}
\altaffiltext{11}{Dipartimento di Fisica ``G. Galilei", Universit\`a di Padova, I-35131 Padova, Italy}
% ...

\begin{document}

\begin{abstract} 
The Extragalactic Background Light (EBL) includes photons with wavelengths from ultraviolet to 
infrared, which are effective at attenuating gamma rays with energy above $\sim 10$ GeV during 
propagation from sources at cosmological distances.
This results in a redshift- and
energy-dependent attenuation of the $\g$-ray flux of extragalactic
sources such as blazars and Gamma-Ray Bursts (GRBs).  The Large Area
Telescope onboard {\em Fermi} detects a sample of
$\g$-ray blazars with redshift up to $z\sim 3$, and GRBs with redshift
up to $z\sim 4.3$.  
Using photons above 10 GeV collected by {\em Fermi} over more than one year of
observations for these sources, we investigate the effect of $\g$-ray flux attenuation by the EBL.  
We place upper limits on the $\g$-ray opacity of the Universe at various
energies and redshifts, and compare this with predictions from
well-known EBL models.  We find that an EBL intensity in the 
optical--ultraviolet wavelengths as great as predicted by the
``baseline'' model of \citet{Stecker06}
can be ruled out with high confidence.
\end{abstract} 

\keywords{diffuse radiation -- dust, extinction -- Gamma rays:
general -- Opacity -- Galaxies: active -- Gamma-ray burst: general}

\section{Introduction}

The {\em Fermi} Gamma Ray Space Telescope was launched 2008 June 11, 
to provide an unprecedented view of the $\g$-ray
Universe. The main instrument onboard {\em Fermi}, the Large Area
Telescope (LAT), offers a broader bandpass \citep[$\sim$ 20~MeV to over
300~GeV;][]{FERMI} and improved sensitivity (by greater than an
order of magnitude) than that of its predecessor instrument EGRET onboard
the {\em Compton Gamma Ray Observatory}~\citep{EGRET}, and the Italian
Space Agency satellite {\em AGILE}~\citep{AGILE}, which was launched in
2007.  The LAT observes the full sky every 3 hr in survey mode leading
to a broadly uniform exposure with less than $\sim 15\%$
variation. The Gamma-ray Burst Monitor, the lower energy ($\sim
8$~keV -- 40~MeV) instrument onboard {\em Fermi}, observes the full
un-occulted sky at all times and provides alerts for transient sources
such as GRBs.

A major science goal of {\em Fermi} is to probe the opacity of the
Universe to high-energy (HE) $\g$-rays as they propagate from their
sources to Earth. Such energetic photons are subject to absorption by
production of electron-positron ($e^-e^+$) pairs while interacting with
low energy cosmic background
photons~\citep{Nishikov61,Gould66,Fazio70} if above the interaction
threshold: $\epsilon_{\rm thr}=(2 m_e c^2)^2/(2E(1-\mu))$ where
$\epsilon$ and $E$ denote the energies of the background photon and
$\g$ ray, respectively, in the comoving frame of the interaction,
$m_ec^2$ is the rest mass electron energy, and $\theta = \arccos(\mu)$
the interaction angle.  Because of the sharply peaked cross section
close to threshold, most interactions are centered around
$\epsilon^*\approx 0.8(E/{\rm TeV})^{-1}$eV for a smooth broadband
spectrum.  Thus, the extragalactic background light (EBL) at UV
through optical wavelengths constitutes the main source of opacity for
$\g$-rays from extragalactic sources (Active Galactic Nuclei: AGN and GRBs) 
in the LAT energy
range.  The effect of absorption of HE $\g$-rays is then reflected in
an energy- and redshift dependent softening of the observed spectrum
from a distant $\g$-ray source.
The observation, or absence, of such
spectral features at HEs, for a source at redshift $z$ can be used to
constrain the $\g\g\to e^+e^-$ pair production optical depth,
$\tau_{\g\g}(E,z)$.

The EBL is dominated by radiation from stars, directly from their surface and via reprocessing by dust in their host galaxies, that accumulated over cosmological evolution. 
Knowledge of its intensity variation with time
would probe models of galaxy and star formation.  The intensity of the EBL
from the near-IR to ultraviolet is thought to be dominated by direct
starlight emission out to large redshifts, and to a lesser extent by
optically bright AGN.  At longer wavelengths the infrared background
is produced by thermal radiation from dust which is heated by
starlight, and also emission from polycyclic aromatic
hydrocarbons~\citep[see e.g.][]{driver08}.

Direct measurements of the EBL is difficult due to contamination by
foreground zodiacal and Galactic light~\citep[e.g.,][]{Hauser01}, and
galaxy counts result in a lower limit since the number of unresolved
sources is unknown~\citep[e.g.,][]{Madau00}.
Furthermore, evolution of the EBL density in the past epochs ($z>0$) that is required to calculate the $\g$-ray flux attenuation from distant sources cannot be addressed by measuring the EBL density at the present epoch (z=0). Hence, several approaches have been developed 
to calculate the EBL
density as a function of redshift. The models encompass different degrees of
complexity, observational constraints and data inputs. Unfortunately, 
the available direct EBL measurements do not constrain these
models strongly at optical-UV wavelengths due to the large scatter in the data points. 
A description
of the different models is beyond the scope of this work; we refer the reader to 
the original works on the various EBL models 
\citep[e.g.,][]{Salamon98, Stecker06, Kneiske02, Kneiske04, Primack05, Gilmore09, Franceschini08, 
Razzaque09, Finke09_model}. We note that all recent EBL models, and in particular all models used in this
paper, use almost identical parameters of a $\Lambda$CDM cosmology model.

For the analyses presented in this work we have made use of the optical depth values $\tau(E,z)$ provided by 
the authors of these EBL models. These models are available via webpages 
\footnote{ \url{http://www.physics.adelaide.edu.au/~tkneiske/Tau_data.html} for Kneiske 2004; \url{http://www.phy.ohiou.edu/~finke/EBL/index.html} for Finke et al. 2010}, analytical 
approximations (as in, e.g., \citet{Stecker06}), published tables (as in, e.g., \citet{Franceschini08}) 
or via private communications (which is the case for, e.g., \citet{Salamon98, Primack05, Gilmore09, Finke09_model} for this work).
Since the optical depth values are usually available in tabular form, for exact values of observed energy $E$ 
and redshift $z$ a linear interpolation of $\tau(E,z)$ is used for arbitrary values of $E$ and $z$ in our 
calculations below.

\begin{figure}
\epsscale{0.9}
\plotone{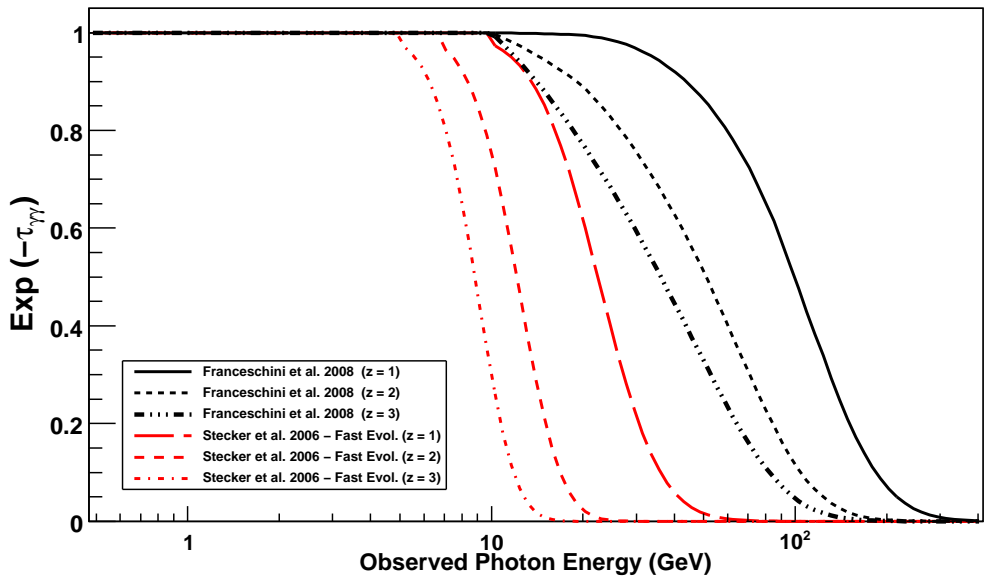}
\caption{Attenuation as a function of observed gamma-ray energy for the EBL models of 
\citep {Franceschini08} and \citep{Stecker06}. These models predict the minimum and maximum absorption 
of all models in the literature, and thus illustrate the range of optical depths predicted in the 
{\em Fermi}-LAT energy range.}
\label{fig:tau_vs_energy}
\end{figure}

The range of predictions by these EBL models is illustrated in Figure~\ref{fig:tau_vs_energy} as a 
function of observed $\g$-ray energy for sources at different redshifts.  
The Universe is optically thin ($\tau_{\g\g} < 1$) to 
$\gamma$-rays with energy below $\simeq 10$ GeV up to redshift $z\simeq 3$, independently of the model 
(see also \citet{dieter}). This is due to 
the rapid extinction of EBL photons shortwards of the Lyman limit. 
Gamma rays below $\sim 10$~GeV are not attenuated substantially because of 
faint far-UV and X-ray diffuse backgrounds.

The primary sources of HE extragalactic $\g$-rays are blazars and gamma ray bursts (GRBs). 
Blazars are active galactic nuclei (AGN) with relativistic plasma outflows (jets) 
directed along our line of sight. GRBs are associated with
the core collapse of massive stars, or might be caused by binary
mergers of neutron stars or neutron star - black hole systems.
Some GRBs produce beamed high-energy radiation similar to the case of blazars but lasting for a short period of time.
GRBs have not been used to constrain EBL absorption during
the pre-{\em Fermi} era mainly because of a lack of sensitivity to transient
objects above 10 GeV. 
The sensitivity of EGRET decreased significantly above 10
GeV, and the field-of-view (FoV) of TeV instruments is small (typically $2-4^\circ$) to
catch the prompt phase where most of the HE emission occurs. The new
energy window ($10- 300$ GeV) accessible by {\em Fermi}, and the wide FoV of the LAT,
makes GRBs interesting targets to constrain EBL absorption in this
energy band.

Evaluating the ratio of the putatively absorbed to unabsorbed fluxes from a
large number of distant blazars and GRBs observed by {\em Fermi} could
result in interesting EBL constraints, as proposed by \citet{chen04},
although intrinsic spectral curvature \citep[e.g.,][]{Massaro06} or
redshift dependent source internal absorption \citep{Reimer07} could make
this, or similar techniques, less effective. \citet{georgan08} have
proposed that Compton scattering of the EBL by the radio lobes of
nearby radio galaxies such as Fornax A could be detectable by the {\em
Fermi}-LAT.  If identified as unambiguously originating from such process, a LAT
detection of Fornax A could constrain the local EBL intensity.

Because the e-folding cutoff energy, $E(\tau_{\g\g}=1)$, from $\g\g$ pair production in
$\g$-ray source spectra decreases with redshift, modern Cherenkov $\g$-ray
telescopes are limited to probing EBL absorption at low redshift due
to their detection energy thresholds typically at or below 50~GeV to 100~GeV \citep{Hinton09}.
Ground-based $\g$-ray telescopes have detected 35 extragalactic
sources to date\footnote{e.g.,\url{http://www.mpi-hd.mpg.de/hfm/HESS/pages/home/sources/}, \url{http://www.mppmu.mpg.de/~rwagner/sources/}}, mostly of the high-synchrotron peaked (HSP) BL Lacertae objects
type.  The most distant sources seen from the ground with a confirmed
redshift are the flat spectrum radio quasar (FSRQ) 3C~279 at $z=0.536$ \citep{Albert08}
and PKS~1510-089 at $z=0.36$ \citep{PKS1510}.  Observations
of the closest sources at multi-TeV energies have been effective in
placing limits on the local EBL at mid-IR wavelengths, while spectra
of more distant sources generally do not extend above 1 TeV, and
therefore probe the optical and near-IR starlight peak of the
intervening EBL \citep[e.g.,][]{Stecker93, Stanev98, Schroedter05,Aharonian99,Aharonian02,Costamante04,
Aharonian06a, Mazin07, Albert08, Krennrich08, Finke09}.

The starting point for constraining the EBL intensity from observations
of TeV $\g$-rays from distant blazars with atmospheric Cherenkov
telescopes is the
assumption of a reasonable intrinsic blazar spectrum, which, in the case of a power law, $dN/dE \propto
E^{-\Gamma_{int}}$ for example, that is not harder than a pre-specified
minimum value, e.g., $\Gamma_{int}\geq\Gamma_{min}=0.67$ or 1.5. Upper limits on the
EBL intensity are obtained when the reconstructed intrinsic spectral
index from the observed spectrum, $\Gamma_{obs}$, presumably softened by EBL absorption of
very high energy (VHE) $\g$-rays, is required to not fall below $\Gamma_{int}$.  The
minimum value of $\G$ has been a matter of much debate, being reasoned
to be $\G_{int}=1.5$ by \citet{Aharonian06a} from simple shock
acceleration theory and from the observed spectral energy distribution (SED) properties of blazars, while 
\citet{stecker07} argued for harder values (less than 1.5)
under specific conditions based on more detailed shock acceleration simulations.
\citet{Katarzynski06} suggested that a spectral index as hard as
$\Gamma_{int}=0.67$ was possible in a single-zone leptonic model if
the underlying electron spectrum responsible for inverse-Compton
emission had a sharp lower-energy cutoff. \citet{boett08} noted that
Compton scattering of the cosmic microwave background radiation by
extended jets could lead to harder observed VHE $\g$-ray spectra, and
\citet{aharonian08} have argued that internal absorption could, in
some cases, lead to harder spectra in the TeV range as well.

A less model dependent approach uses the (unabsorbed) photon index as
measured in the sub-GeV range as the intrinsic spectral slope
at GeV-TeV energies. This method has recently been applied
to PG~1553+113 \citep{lat1553} and 1ES~1424+240 \citep{lat1424,prandini10} to derive upper
limits on their uncertain redshifts, and to search for EBL-induced spectral softening
in {\em Fermi} observations of a sample of TeV-selected AGN \citep{TeVselected}. 

Attenuation in
the spectra of higher redshift objects ($z \gtrsim 1$) may be 
detectable at the lower
energies that are accessible to the {\em Fermi}-LAT, i.e., at $E\approx
10-300$ GeV. Gamma rays at these energies are attenuated mainly
by the evolving UV background, which is produced primarily by young
stellar populations and closely traces the global star-formation rate.
Observations with {\em Fermi} of sources out to high redshift could
therefore reveal information about the star-formation history of the
Universe, as well as the uncertain attenuation of UV starlight by
dust.

In this paper we present constraints on the EBL intensity of the Universe derived from {\em Fermi}-LAT observations of blazars and GRBs.
The highest-energy $\g$-rays
from high redshift sources are the most effective probe of
the EBL intensity, and consequently a powerful tool for investigating
possible signatures of EBL absorption.  In contrast to ground-based
$\g$-ray detectors, {\em Fermi} offers the possibility of probing the
EBL at high redshifts by the detection of AGN at $\gtrsim 10$ GeV
energies out to $z>3$, and additionally by the detection of GRB
080916C at a redshift of $\sim 4.35$ \citep{abdo09,Greiner09}. GRBs
are known to exist at even higher redshifts \citep[GRB 090423 is the current
record holder with $z\sim$ 8.2]{Tanvir09}. Therefore
observations of these sources with {\em Fermi} are promising
candidates for probing the optical-UV EBL at high redshifts that are
not currently accessible to ground-based (Cherenkov) telescopes.

In Section~\ref{sec:analysis} we describe our data selections,
the {\em Fermi} LAT AGN and GRB observations during the first year of
operation and analysis, and we discuss potential biases in the selection.  Our
methodology and results are presented in Section~\ref{sec:methods}.  We
discuss implications of our results in Section~\ref{sec:discussion}, and
conclude in Section~\ref{sec:conclusion}.

In the following, energies are in the observer frame except where noted otherwise.

\section{Observations and data selection} 
\label{sec:analysis}

The {\em Fermi} LAT is a pair-conversion detector sensitive to $\g$-rays
with energies greater than 20 MeV.  The LAT has a peak effective area
$\geq 8000$ cm$^2$ at energies greater than 1 GeV relevant for most of the event
selections considered in this analysis and a large, $\sim2.4$ sr,
field-of-view.  The angular resolution for the 68\%
containment radius is $\sim0.6^{\circ}$ for 1 GeV photons that convert
in the upper layers  of the tracker ({\it front} events) and about a factor of 2 larger for those that
convert in the bottom layers of the tracker ({\it back} events).  A simple acceptance-averaged approximation for
the $68\%$ containment angle that is helpful to illustrate the energy
dependent PSF is $\langle \theta_{68}(E) \rangle = (0.8^\circ) \times
(E/{\rm GeV})^{-0.8} \bigoplus (0.07^\circ)$.  A full description of
the LAT instrument is reported in \citet{FERMI}.

The data set used for the analysis of the AGNs includes
LAT events with energy above $100$ MeV that were
collected between 2008 August 4 and 2009 July 4.  LAT-detected GRBs are considered up to 2009 September
30.  A zenith angle cut of $105^\circ$ was applied in
order to greatly reduce any contamination from the Earth albedo.
Blazars and GRBs have different emission characteristics,
which result in different analysis procedures here. The event rate detected by
the LAT in the vicinity (68\% confidence radius) of a blazar is largely background dominated and only continuous
observations over long time scales allow the detection of the underlying blazar emission.
To minimize the background contamination when analyzing blazar data we use the
``diffuse'' class events,
which provide the purest $\g$-ray sample and
the best angular resolution.  
GRBs, on the other hand, emit most of their radiative $\g$-ray power on very short time scales (typically on the order of seconds) 
where the event rate can be considered mostly
background free (at least during the prompt emission of bright
bursts).  It is therefore possible to loosen the event class selection
to increase the effective area at the expense of a higher background
rate which is still small on short time scales for bright bursts.  The
``transient'' class was designed for this specific purpose and we use
these events for GRB analysis.\footnote{see \cite{FERMI} for further details on LAT event selection.}

\subsection{AGN sample and potential biases} 
\label{sec:agn_sample}

We use blazars extracted from the First LAT AGN Catalog
\citep[1LAC;][]{1LAC} as the AGN source sample to probe the UV
through optical EBL.  This catalog contains 671 sources at high Galactic latitude
($|b|>10^\circ$) associated with high-confidence with blazars and
other AGNs that were detected with a Test Statistic\footnote{The test statistic (TS) is defined as
$TS = -2 \times (log(L_{0})-log(L_{1}))$ with $L_0$ the likelihood of the
Null-hypothesis model as compared to the likelihood of a
competitive model, $L_1$, (see Section 3.2.2).}
$TS>25$ during the first 11 months of
science operation of the LAT. Detection of correlated
multiwavelength variability was required in order to establish a
source as being identified.

Source associations were made with a
Bayesian approach \citep[similar to][]{mattox01}. 
The Bayesian approach for source association
implemented in the {\it gtsrcid} tool of the LAT {\it ScienceTools}
package\footnote{http://fermi.gsfc.nasa.gov/ssc/data/analysis/scitools/overview.html} uses only spatial coincidences between LAT
and the counterpart sources. 
Candidate source catalogs used for this procedure include CRATES
\citep{healey07}, CGRaBS \citep{healey08} and the Roma-BZCAT
\citep{Massaro09}, which also provide optical classifications and
the latter two provide also spectroscopic redshifts for the sources.  
See \citet{1LAC} for further
details on the source detection and association algorithims refered to
here.

As discussed below, some methods applied here require one to distinguish
among the different blazar source classes. Flat-spectrum radio quasars (FSRQs) 
and BL Lacs are discerned by their observed optical emission line equivalent widths
and the Ca II break ratio \citep[e.g.,][]{stocke91,mar96} following
the procedure outlined in \citet{LBAS}. The BL Lac class itself is
sub-divided into Low-, Intermediate-, and High-Synchrotron peaked BL
Lacs (denoted as LSP-BLs, ISP-BLs and HSP-BLs, respectively) by
estimating the position of their synchrotron peak, $\nu_{\rm peak}^s$,
from the indices $\alpha_{ox}\simeq 0.384\cdot\log(f_{\rm
5000A}/f_{\rm 1keV})$ and $\alpha_{ro}\simeq 0.197\cdot\log(f_{\rm
5GHz}/f_{\rm 5000A})$ determined by the (rest frame) optical $(f_{\rm
5000A})$, X-ray $(f_{\rm 1keV})$ and radio $(f_{\rm 5GHz})$ flux
densities listed in the online version of the Roma-BZCAT blazar
catalog \citep{Massaro09}, and using an empirical relationship
between those broadband indices and $\nu_{\rm peak}^s$ as derived in
\citet{SEDpaper}. LSP-BLs have their synchrotron peak at $\nu_{\rm
peak}^s < 10^{14}$~Hz, ISP-BLs at $10^{14}$~Hz$\leq \nu_{\rm peak}^s \leq
10^{15}$~Hz and HSP-BLs at $\nu_{\rm peak}^s > 10^{15}$~Hz.  This is
found to be in agreement with the classifications used in
\citet{spec_paper,SEDpaper}.  Nearly all the 296 FSRQs are of
LSP-type, while only 23\% of the 300 BL Lacs are LSP-BLs, 15\% are
ISP-BLs, and 39\% are HSP-BLs, 72 AGNs could not be classified, and 41
AGNs are of other type than listed above.

Redshift information on the sources is extracted from the counterpart
source catalogs (CRATES, CGRaBS, Roma-BZCAT).  While all the redshifts
of the 1LAC FSRQs are known, only 42\% of the high-confidence BL Lacs have measured redshifts. 
Obviously, AGN without redshift information are not used in the present work.

The intrinsic average photon indices of {\em Fermi} blazars in the LAT energy range indicate a
systematic hardening with source type from $\sim 2.5$ for FSRQs via
$\sim 2.2$ for LSP- and $\sim 2.1$ for ISP-, to $\sim 1.9$ for HSP-BLs
\citep{spec_paper}.  On the other hand, their redshift distributions
systematically decrease from the high-redshift (up to $z\sim 3.1$)
FSRQs, via LSP-BLs located up to a redshift $z\sim 1.5$, down to the
mostly nearby HSP-BLs at $z<0.5$.  This mimics a spectral softening
with redshift if blazars are not treated separately by source type. A search
for any systematic spectral evolution must therefore differentiate
between the various AGN sub-classes (see below).

To detect absorption features in the HE spectra ($>10$~GeV; see Fig.~\ref{fig:tau_vs_energy}) of {\em Fermi} blazars, 
a thorough understanding of their intrinsic spectra,
including variability, and source internal spectral features is required.  Most
blazars do not show strong spectral variability in the LAT energy
range on $\gtrsim$ week scales \citep{spec_paper}, despite often strong
flux variability \citep{variability_paper}.  Indeed at least three blazars
which turn out to constrain the UV EBL the most (see
Section~\ref{sec:HEP method}, \ref{sec:likelihood}), show a $>99\%$
probability of being variable in flux (using a $\chi^2$ test) with a
normalized excess variance of $\sim 0.02-0.2$ on time scales of hours to weeks.  
PKS~1502+106 (J1504+1029) is one of
the most constraining sources in the sample. It displayed an exceptional flare in
August 2008 with a factor $\sim 3$ increase in flux
within $\sim 12$~hours  \citep{lat1502}.  During this flare a flatter (when brighter)
spectral shape was evident. The spectral curvature at the high energy end increased 
with decreasing flux level.
  If the high energy ($\gtrsim
10$~GeV) photons are emitted during such flare activity, the
constraints on the $\g$-ray optical depth would be tighter if only the 
flare-state spectral data were used.  Because of limited photon statistics
during the flare, however, we use the more conservative time averaged
spectrum in the present analysis.

Absorption in radiation fields internal to the source (e.g., accretion
disk radiation, photon emission from the broad line region) may cause
a systematic break in the $\g$-ray spectra that coincidentally mimics
EBL attenuation \citep{Reimer07}.  In the case where such internal absorption
occurs, its redshift dependence is guaranteed, even in the absence
of accretion evolution.  This is because of the redshifting of that
energy where the interaction probability is maximum
\citep{Reimer07}.  Any technique that explores systematic variaton of
observables (e.g., changes in spectral slope, flux ratios, e-folding
cutoff energy) with redshift to single out EBL-induced absorption
features in blazars with luminous accretion disk radiation (possibly
indicated by strong emission lines) might therefore suffer from such
a bias.

All bright strong-line {\em Fermi} blazars (i.e., {\em Fermi}-FSRQs and some LSP-
and ISP-BLs), however, have been found to show spectral breaks already
in the 1-10 GeV (source frame) range \citep{spec_paper}. This is too low in
energy to be caused by EBL-attenuation for their redshift range $\lesssim 3$ 
(see Fig.~\ref{fig:tau_vs_energy}).  
Although it is not clear if these breaks are due to internal absorption, 
the spectral softening results
in low photon counts at energies $\gtrsim 10$~GeV where EBL
absorption is expected.  Spectra of all bright HSP-BLs and some
ISP-BLs, on the other hand, can be well represented by simple power
laws without any signs of curvature or breaks.  
This indicates not only do they
not have significant internal absorption in the $\gamma$-ray band, but
also the absence of significant EBL absorption, which is expected to be
beyond the LAT energy range for this nearby ($z\lesssim 0.5$) blazar
population.

Consequently, as we show in Section~\ref{sec:flux ratios}, it remains challenging to quantify EBL absorption effects
in the LAT energy range based on population studies.  On the other hand, the determination of
the EBL-caused absorption features from individual blazars requires
bright, high redshift objects with spectra extending to $\gg10$GeV (Fig.~\ref{fig:tau_vs_energy}), 
and we focus on these blazars in
Section~\ref{sec:individual sources}.

\subsection{GRB sample and potential biases}
\label{sec:grb_sample}

The {\em Fermi} LAT has detected 11 GRBs from the beginning of its science
operation (August, $4^{th}$ 2008) until 30 September 2009,
6 of which have redshift measurements. Figure \ref{fig:emax_vs_redshift}
shows the redshift and highest energy event associated with each of these
GRBs.  The
probability of non-association is extremely small (see Table~1).

GRB prompt emission is highly variable and shows signs of spectral
evolution, a source of systematics to be considered carefully.  Our
approach in this paper is to restrict ourselves to the analysis of small
time windows during the GRB emission where the temporal behavior does
not seem to change significantly.

The GRB spectral behavior is well-represented by the Band function
(Band et al. 1993) in the keV--MeV range.  An additional hard, $\G
\sim 1.5$--2, power-law component, dominating at $\gtrsim 100$~MeV,
has now been firmly identified in a few GRBs:
GRB090510 \citep{grb090510}, GRB090902B, \citep{grb090902B}, GRB090926A \citep{grb090926A}.
Its absence in other LAT
bursts could well be due to limited photon statistics.  We assume that
the power-law component extends well beyond 100~MeV up to $\sim
10$~GeV, below which EBL absorption is negligible (see Fig.~\ref{fig:tau_vs_energy}).  EBL
absorption is then expected to soften the power-law spectra from the
extrapolation of the intrinsic/unabsorbed spectra beyond $\sim
10$~GeV.  Systematic effects will, of course, occur when an intrinsic
spectrum at high energies differs from this extrapolation.  Source
internal and/or intrinsic absorption via pair creation, e.g., would
produce a curvature of the spectrum at higher energy which could be
misinterpreted as an EBL absorption effect.  
Such a spectral break, which could be due to intrinsic pair creation, was detected 
in the LAT data from GRB 090926A \citep{grb090926A} but we note that a corresponding 
roll-off in the intrinsic spectrum can only make our limits on the 
$\g$-ray optical depth more conservative.
By contrast, a rising spectral component above $>$10 GeV would make our limits less constraining, but in the absence of any evidence for inverted gamma-ray spectra in GRBs, we consider this possibility unlikely.

\section{Analysis of $\g$-ray flux attenuation and results} 
\label{sec:methods}

Assuming that high-energy photon absorption by the EBL is the sole
mechanism that affects the $\g$-ray flux from a source at redshift
$z$, the observed (i.e. absorbed) and unabsorbed fluxes at the observed energy $E$ 
can be related by the optical depth,
$\tau_{\g\g}(E,z)$, as
\be
F_{obs}(E) = \exp[-\tau_{\g\g}(E,z)]F_{unabs}(E).
\label{flux_relation}
\ee
This is the primary expression that we use to (i) explore $\g$-ray
flux attenuation in the EBL from AGNs by means of
a redshift-dependent flux ratio between a low- and a high- energy band; 
(ii) constrain EBL
models which predict $\tau_{\g\g}(E,z)$ values much higher than the
optical depth that would give the observed fluxes from individual blazars
and GRBs; and (iii) put upper limits on the $\g$-ray optical depth calculated
from the observed flux of individual blazars and GRBs, and the extrapolation
of the unabsorbed flux to high energies.  We discuss these methods
and the results from our analysis below.

\subsection{Flux ratios - a population based method}
\label{sec:flux ratios}

Because of inherent uncertainties in the determination of the intrinsic
spectrum ($\G_{int}$) for any given blazar in the pre-{\em Fermi} era,
\citet{chen04} proposed the average ratio $F(>10~{\rm GeV})/F(>1~{\rm
GeV})$ for all blazars with significant detections above 1 GeV, weighted
according to the errors in $F(>1~{\rm GeV})$, as a
redshift-dependent tracer of the EBL attenuation of $\g$-ray flux.
The average flux ratio could then be compared with the predictions of
the EBL models, taking selection effects into account.  This approach
assumes that the blazars are sampled from a homogeneous distribution
with a single redshift-dependent luminosity function and a single
intrinsic spectral index distribution.  Preliminary results from 
{\em Fermi} \citep{LBAS} indicate that this assumption is inadequate.
Consequently, we have calculated the average flux ratios for the
different classes of blazars and discuss the results below.

Among the AGN sample described in Section~\ref{sec:agn_sample} we find
that 237 FSRQs, 110 BL Lacs and 25 other AGNs are clean\footnote{i.e., its 
association probability is at least 80\%, it is the sole AGN associated with the 
corresponding $\gamma$-ray source, 
and it is not flagged to have problems that cast doubt on its detection \citep{1LAC}} 
1LAC associations 
with known redshift and detectable fluxes at energies
$\ge 1$~GeV. There are 30 LSP-, 18 ISP- and 60 HSP- BL Lacs in
this sub-sample. 

Of these AGN, only 22 FSRQs, 49 BL Lacs, and 1 other AGN have flux detections
rather than upper limits above 10~GeV, including 10 LSP-, 6 ISP-, and 33 HSP-BL Lacs.
For each of these BL Lacs and FSRQs, we calculated the ratio
between the fluxes above 10~GeV and 1~GeV and their corresponding
statistical errors following \citet{chen04}.

\begin{figure}
\epsscale{0.7}
\plotone{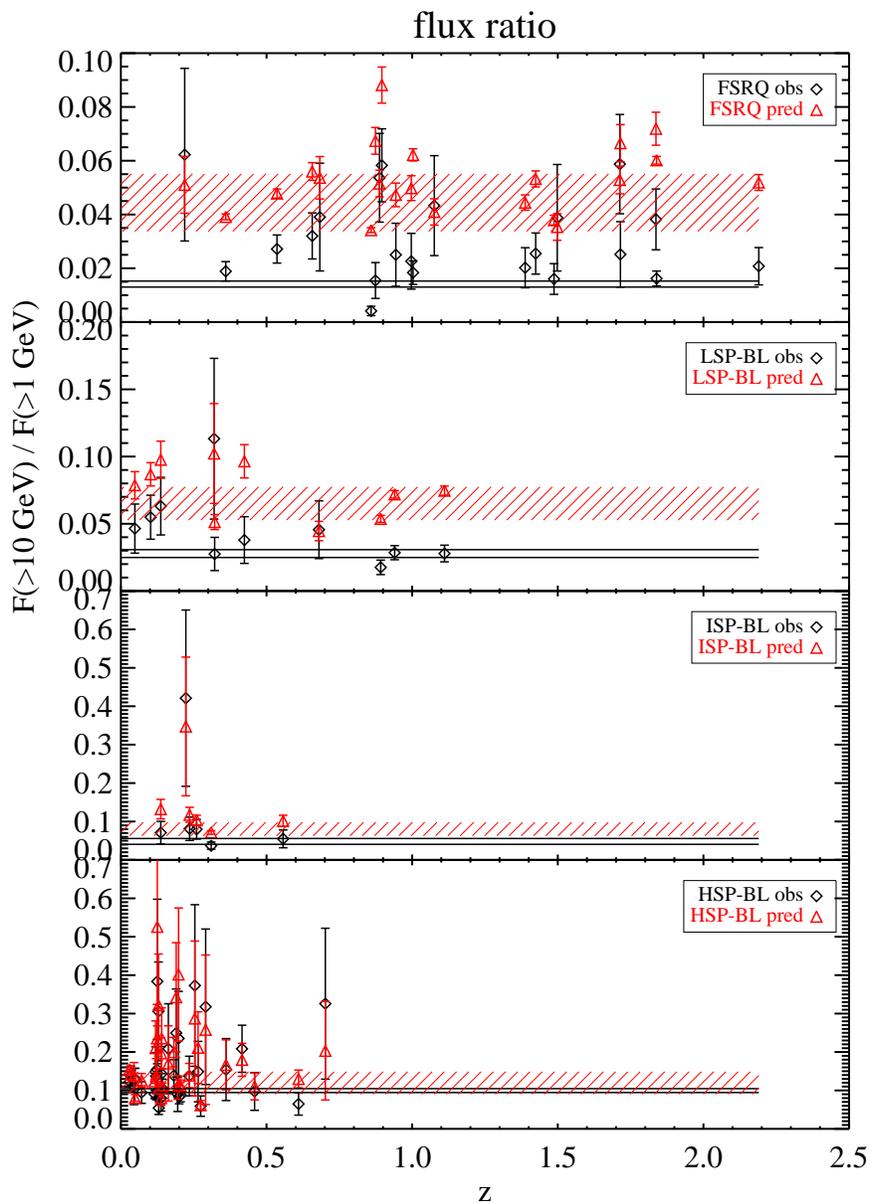}
\caption{Flux ratio $F(\geq 10~{\rm GeV}) / F(\geq 1~{\rm GeV})$ as a
function of redshift, in the {\em Fermi} LAT energy range, for FSRQs and BL
Lac populations.  The black diamonds are the observed ratios, while the red
triangles show the ratio expected assuming an unbroken power law and no
EBL attenuation. The black horizontal solid lines and red cross-hatched regions
correspond to the mean observed ratios and expected ratios with errors.}
\label{fig:ratio}
\end{figure}

\begin{table}[htdp]
\begin{center}
\begin{tabular}{|c|c|c|c|c|c|c|}
\hline
Blazar type & Num & $\Gamma$ & ratio (pred) & mean ratio (obs) & red. $\chi^2$ & prob\\
\hline\hline
FSRQ & 22 & $2.3 \pm 0.1$ & $0.04 \pm 0.01$ & $0.014 \pm 0.001$ & 4.38 & $1.8\times 10^{-10}$ \\
LSP-BL & 10 & $2.2 \pm 0.1$ & $0.07 \pm 0.01$ & $0.028 \pm 0.003$ & 1.65 & 0.11\\
ISP-BL & 6 & $2.1 \pm 0.1$ & $0.08 \pm 0.02$ & $0.048 \pm 0.008$ & 1.86 & 0.11\\
HSP-BL & 33 & $1.9 \pm 0.1$ & $0.12 \pm 0.03$ & $0.100 \pm 0.005$ & 1.29 & 0.13\\
\hline
\end{tabular}
\end{center}
\caption{Spectral indices, mean predicted and observed flux ratios, and reduced $\chi^2$ and probability for blazar sub-populations}
\label{table:ratio}
\end{table}%

Figure~\ref{fig:ratio} shows the observed flux ratios for the FSRQ population and
BL Lac sub-populations as well as the ratios predicted according to the
1FGL spectral index of each blazar, assuming an unbroken power law and
no EBL attenuation. Table~\ref{table:ratio} shows the mean spectral index, mean
flux ratios observed and expected, and the reduced $\chi^2$ and associated probability
given a parent distribution with constant flux ratio. As the blazar classes progress from
FSRQ through LSP-BL, ISP-BL, and HSP-BL, 
\begin{enumerate}
\item the range of redshifts becomes narrower;
\item on average, the spectra become harder;
\item both the predicted and observed mean flux ratios increase;
\item the difference between the predicted and observed flux ratios decreases.
\end{enumerate}

The trend in the predicted flux ratios is a direct consequence of the hardening of the spectra
as a function of source class, while
the difference between the predicted and observed flux ratios is due to the fact that the curvature
of the spectra decrease as the HE peak of the SED moves through the Fermi-LAT energies.
The apparent discrepancies between the flux ratios for different
blazar sub-populations arise from the fact that the LAT samples
different parts of the blazar SED for these sub-classes.  Indeed, a
redshift distribution of the flux ratios for the combined blazar
populations would show a strong, apparently decreasing trend, giving the
appearance of an EBL absorption effect. When we separate the blazars into 
sub-populations, we find no significant redshift dependence of the flux ratios 
within each sub-population. The dearth of sources at high redshift
and the large spread of spectral indices make it difficult to use the
mean trend in the flux ratio as a function of redshift. To set
upper limits on the $\gamma$-ray optical depth, we need to rely on the 
spectra of individual blazars, despite the increased dependence on
the blazar emission model this entails.

The flux ratio versus redshift relationship for BL Lacs is therefore
primarily due to the differing intrinsic spectral characteristics of
BL Lacs, rather than from EBL absorption.  This test is a reminder of
the importance of a careful consideration of the intrinsic spectral
characteristics of the source populations chosen to probe EBL
absorption.

\subsection{Constraints on EBL models from individual source spectra}
\label{sec:individual sources}

The sensitivity of the LAT over a broad energy range provides a unique
opportunity to probe $\g$-ray spectra from AGNs and GRBs at $< 10$~GeV
where EBL absorption is negligible and at $\gtrsim 10$~GeV where EBL
absorption can be substantial (see Fig.~\ref{fig:tau_vs_energy}).  Thus extrapolations of the unabsorbed
flux at low energies from individual sources to high energies, and
assuming that the intrinsic spectra do not become harder at high
energies, allows us to derive a measure of the total absorption
(source in-situ and in EBL).  We note that this is the only assumption
made for the following methods.
Furthermore, since any intrinsic
spectral curvature or internal absorption effects can not be decoupled
from EBL-caused curvature, the constraints derived below shall be
considered as conservative upper limits on the EBL-caused opacity.
These are then confronted with various EBL models. Clearly, high EBL
density models possess a higher probability of being constrained by these
methods than low density ones. In the following, we use two methods:
the highest energy photon (Section~\ref{sec:HEP method}) and the
likelihood (Section~\ref{sec:likelihood}) methods.

\begin{figure}
\epsscale{1.0}
\centering
  \begin{minipage}[b]{15cm} 
     \includegraphics[width=12cm]{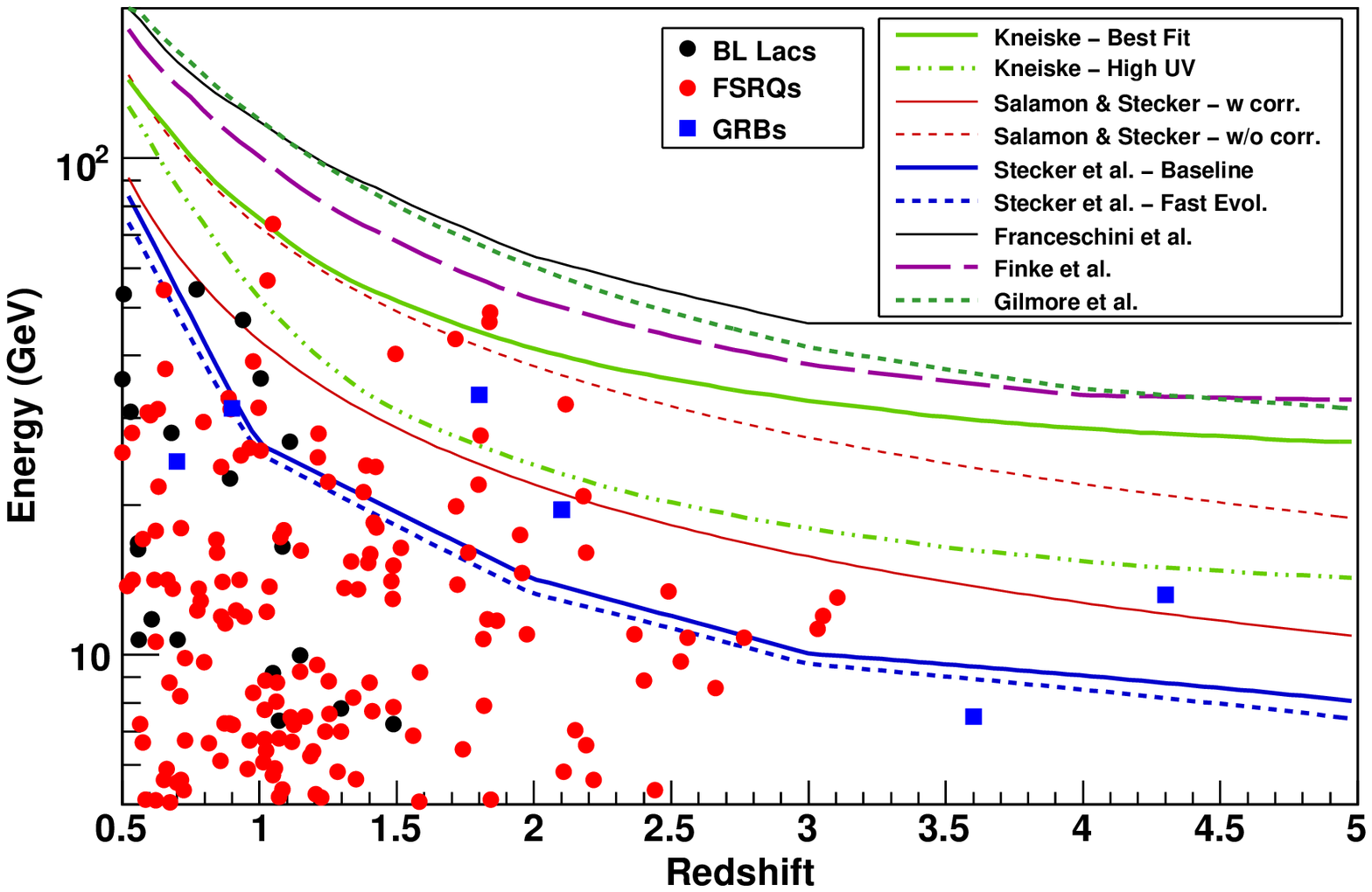}
  \end{minipage}
  \begin{minipage}[b]{15cm}
     \includegraphics[width=12cm]{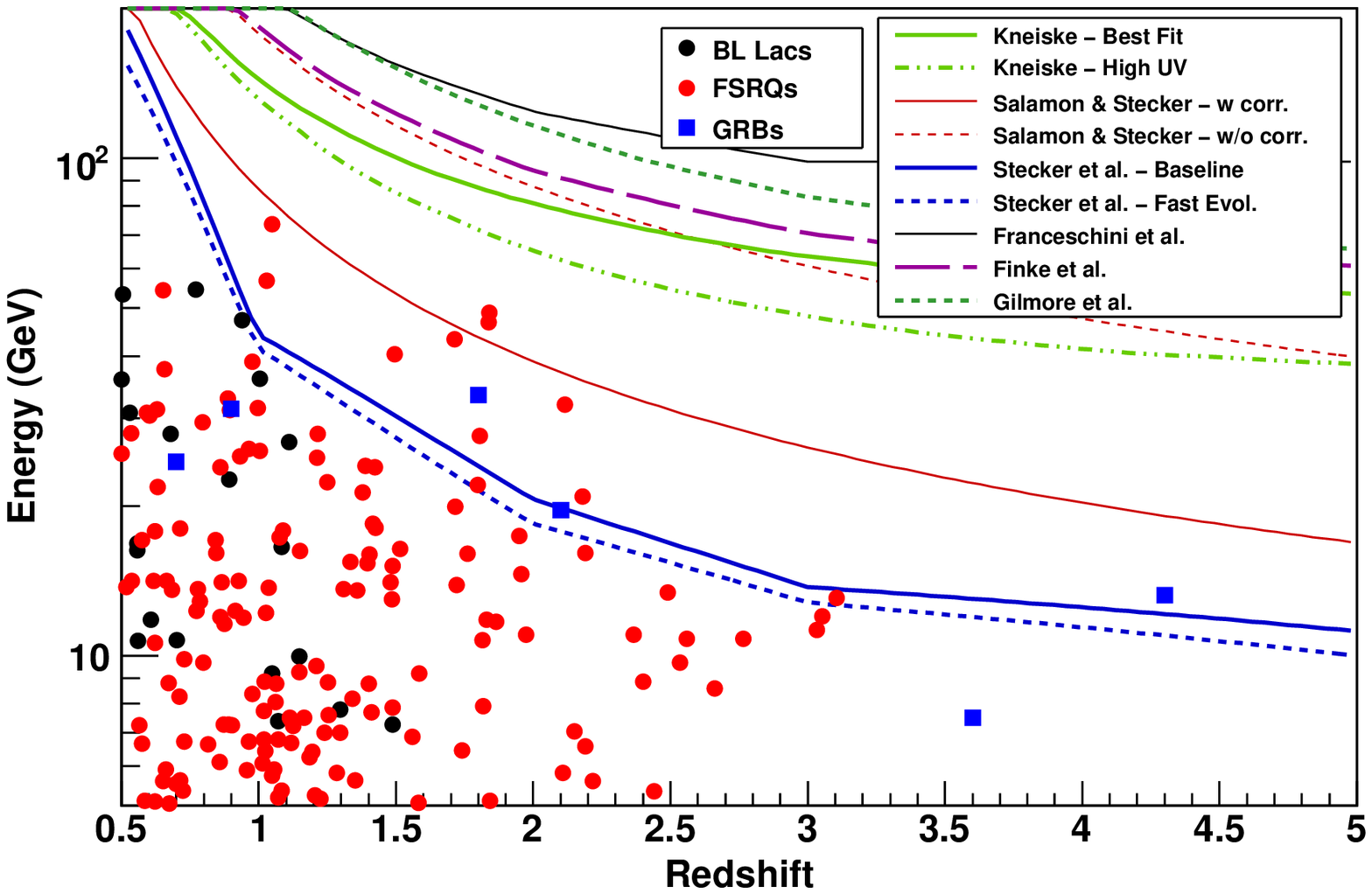} 
  \end{minipage}
\caption{Highest-energy photons from blazars and GRBs from different
redshifts.  Predictions of $\g\g$ optical depth $\tau_{\g\g} = 1$ (top
panel) and $\tau_{\g\g} = 3$ (bottom panel) from various EBL models
are indicated by  lines. Photons above model predictions in this figure traverse an EBL medium with a high $\gamma$-ray opacity. The likelihood of detecting such photon considering the spectral characteristics of the source are considered in the method presented in section \ref{sec:HEP method}.}
\label{fig:emax_vs_redshift}
\end{figure}

\begin{table}
\begin{scriptsize}
\begin{center}
\renewcommand{\arraystretch}{1.2}
\begin{tabular}{|c|c|c|c|c|c|c|c|} 
\hline
Source & $z$ & $E_{max}$ (GeV) & Conv.\ Type & $\Delta E/E$  &  68\% radius &  Separation & Chance Probability \\
\hline\hline
J1147-3812 & 1.05 & 73.7 & {\em front} & 10.7 \% & $0.054^\circ$ & $0.020^\circ$ & $7.0\times10^{-4}$ \\
(PKS 1144-379) & & & & &\\
\hline
J1504+1029 & 1.84 & 48.9 & {\em back} & 5.4\% & $0.114^\circ$  & $0.087^\circ$ & $5.6\times10^{-3}$ \\
(PKS 1502+106) &      & 35.1 & {\em back} & 12.4\% & $0.117^\circ$ & $0.086^\circ$& $9.8\times10^{-3}$\\
         &      & 23.2 & {\em front} & 7.2\% & $0.072^\circ$ &  $0.052^\circ$& $5.6\times10^{-3}$\\
\hline
J0808-0751 & 1.84 & 46.8 & {\em front} & 9.7\% & $0.057^\circ$ & $0.020^\circ$ & $1.5\times10^{-3}$ \\
(PKS 0805-07)&      & 33.1 & {\em front} & 5.9\% & $0.063^\circ$ & $0.038^\circ$ &$2.7\times10^{-3}$\\
         &      & 20.6 & {\em front} & 8.9\% & $0.075^\circ$ & $0.029^\circ$ &$6.9\times10^{-3}$\\
\hline
J1016+0513 & 1.71 & 43.3 & {\em front} & 11.4\% & $0.054^\circ$ & $0.017^\circ$ & $1.2\times10^{-3}$ \\
(CRATES J1016+0513) &      & 16.8 & {\em front} & 6.3\%  & $0.087^\circ$ & $0.035^\circ$ &$8.2\times10^{-3}$\\
         &      & 16.1 & {\em front} & 7.6\% & $0.084^\circ$ & $0.018^\circ$ &$8.2\times10^{-3}$\\
\hline
J0229-3643 & 2.11 & 31.9 & {\em front} & 10.7\% & $0.060^\circ$ & $0.035^\circ$ & $1.7\times10^{-3}$ \\
(PKS 0227-369) & & & & &\\
\hline
GRB 090902B & 1.82 & 33.4 & {\em back} & 10.5\% &  $0.117^\circ$ &  $0.077^\circ$ & $6.0\times 10^{-8}$ \\
\hline
GRB 080916C & 4.24 & 13.2 & {\em back} & 11.6\% &  $0.175^\circ$ & $0.087^\circ$ & $2.0\times 10^{-6}$ \\
\hline
\end{tabular}
\end{center}
\end{scriptsize}
\caption{List of blazars and GRBs detected by the LAT which have
redshift measurements, and which constrain the EBL opacity the most. For each source, J2000 coordinate based name (other name), the energy of the 
highest-energy photon (HEP), the conversion type of the event ({\em front} or {\em back}) of the instrument, the energy 
resolution, $\Delta E/E$, for 68\% containment of the reconstructed incoming photon energy, and the 68\%
containment radius based on the energy and incoming direction in instrument coordinates of the event, the separation from the source and the chance probability of the HEP being from
the galactic diffuse or isotropic backgrounds 
are also listed. The energy resolution for the GRB HEP events is taken from \citet{grb090902B} and \citet{grb080916C} using
the respective lower energy bounds.
The three highest-energy photons are listed for those sources that have multiple constraining photons.}
\label{high-energy-photons}
\end{table}

\subsubsection{Highest energy photons}
\label{sec:HEP method}

A simple method to constrain a given EBL model is to calculate the
chance probability of detecting a photon with energy $E \geq E_{max}$, where
$E_{max}$ is the energy of the most energetic photon that we would expect when the source intrinsic spectrum is folded with the
optical depth from the specific EBL model we want to test. We derive a conservative estimate of the
 intrinsic flux of the source by
extrapolating the unabsorbed spectrum at low energies to high
energies.  We consider the LAT spectrum to be representative of the
intrinsic spectrum at energies where the EBL is supposed to absorb
less than $\sim 1 \%$ of the photons for the most opaque models. This
corresponds to an energy of around $10$~GeV  
(down to $\sim 6$ GeV for GRB 080916C at $z \sim 4.3$). 
Best fit spectral
parameters of this ``low-energy'' unabsorbed spectrum were derived for
all sources of the HEP set (see Table~\ref{tab:fit}).  The spectrum is
assumed to be a power law unless a significant deviation from this
shape is measured at $\lesssim 10$~GeV  (as is indeed observed from, e.g.,
FSRQs at GeV energies). This is the case for source J1504+1029, for which a log parabola 
model provides the best fit.

Iterating through the source list described in
Section~\ref{sec:agn_sample} and Section~\ref{sec:grb_sample} we find the
energy $E_{\rm max}$ of the highest-energy photon detected within the 68\%
containment radius (using the specific P6\_V3\_DIFFUSE instrument response
functions for {\em front} and {\em back} events) of each source position.  The resulting $E_{\rm max}$
versus source redshift is shown in Figure~\ref{fig:emax_vs_redshift}
for sources with $z>0.5$, and compared to the energy at which the optical depth $\tau_{\gamma\gamma}$ is 
equal to 1 and 3 according to the various EBL models. As shown in this figure, 5 AGN have $E_{\rm max}$ that is 
significantly greater $(\gtrsim 2)$ than the energy at which $\tau_{\gamma\gamma}=3$ for the 
``baseline EBL model'' of \citet{Stecker06}. These 5 AGN (and 2 comparable GRBs) have emitted a number 
of events (hereafter {\em highest-energy photons} or HEP) that populate a region of the $E_{max}-z$ phase
space where EBL attenuation effects are predicted to be
significant. These (henceforth called ``HEP set'') will be used in the following sections to 
constrain EBL models and to calculate the
maximum amount of EBL attenuation that is consistent with the LAT
observations\footnote{Only the highest energy photon from each source is shown in Figure~\ref{fig:emax_vs_redshift}.  
There are a few sources however with more than one constraining photon as indicated in Table~\ref{high-energy-photons}.}.

It is possible that the high-energy  photons considered here may not be emitted in the high-redshift source and instead are originated in any of the following background sources:  Galactic $\gamma$-ray diffuse,  isotropic (Extragalactic $\gamma$-ray plus charged-particle residuals) or  a nearby point source.  The likelihood of detecting any of these background photons within the 68\% containment radius used to find the HEP set  is quantified by computing the number of expected events within the 68\% containment radius at the location of the source as determined by the best fit background model (Galactic and isotropic diffuse + point sources) and the instrument acceptance.  The last column of Table~\ref{high-energy-photons} shows such probability for  photons in the HEP set. These chance probabilities, although being fairly small, are non-negligible (at least in the case of blazars) if one would like to set significant constraints on specific EBL models by using this HEP. We later describe how this probability for the HEP to be a background fluctuation was incorporated in our final results for this method. For now we will assume that this HEP is indeed from the source and we will first derive the type of constraints it allows us to set on different EBL models.
We also note that a stricter set of cuts ({\em extradiffuse}) has been
developed by the LAT team to study the Extragalactic $\g$-ray
background \citep{EGBpaper}. Despite the decreased $\g$-ray acceptance
we find all photons in the HEP set to be retained when using these
selection cuts.

Monte-Carlo simulations are performed in order to test a particular EBL model with the derived intrinsic
spectrum absorbed by the EBL as the Null-hypothesis.   The simulations were performed using {\em gtobssim}, one of the science tools distributed by the {\em Fermi} science support center and the LAT instrument team. For each simulation we define the  unabsorbed spectrum of the source as a power law (or log parabola in the case of  J1504+1029) with spectral parameters drawn randomly from the best-fit values (and corresponding uncertainty)  shown in Table~\ref{tab:fit}. EBL absorption is applied according to the optical depth values of the considered model, and finally, the position and orientation of the {\em Fermi} satellite  during the time interval described in Section~2  is used to account for the instrument acceptance that corresponds to the observations. The highest-energy photon from the simulated data is obtained following the exact same cuts and analysis procedure that was used for the data. 

\begin{table}
\begin{center}
\renewcommand{\arraystretch}{1.2}
\begin{tabular}{|c|c|c|c|c|c|c|cl} 
\hline
Source & normalization $N_0$& photon index $\Gamma$  & $\Gamma$, $\beta$, $E_b$/GeV & TS \\
& ($ 10^{-7} \mbox{ph cm$^{-2}$ s$^{-1}$ MeV$^{-1}$}$)& (PL) & (LP) & \\
\hline\hline
J1147-3812 & $0.570 \pm 0.081$ & $2.38 \pm 0.09$ & ... & 221 \\
\hline
J1504+1029 & $(1.84 \pm 0.23)\cdot10^{-4}$ & ... & $2.36 \pm 0.03,$ & 34638\\
&&& $0.09 \pm 0.01,$ & \\ 
&&& $2.0 \pm 0.1$& \\
\hline
J0808-0751 & $1.212 \pm 0.078$ & $2.09 \pm 0.04$ & ... & 1498 \\
\hline
J1016+0513 & $1.183 \pm 0.078$ & $2.27 \pm 0.05$ & ... & 1220 \\
\hline
J0229-3643 & $0.789 \pm 0.075$ & $2.56 \pm 0.07$ & ...& 394\\
\hline
J1012+2439 & $0.552 \pm 0.058$ & $2.21 \pm 0.07$ & ... & 443\\
\hline
GRB 090902B  & $146 \pm 56$ & $1.40 \pm 0.37 $ &... & 1956 \\
\hline
GRB 080916C  & $1146 \pm 199$ & $2.15 \pm 0.22 $ &... & 1398 \\
\hline
\end{tabular}
\caption{Parameter values of the power law (PL) fits $dN/dE = N_0(-\Gamma+1) E^{-\Gamma}/[E_{max}^{-\Gamma+1}-E_{min}^{-\Gamma+1}]$ 
in the range $E_{\rm min}=100$~MeV to $E_{\rm max}=10$~GeV of the sources (AGN and GRBs) listed in
Table~\ref{tab:spectres} except for source J1504+1029 where a log parabolic parametrization (LP) 
$dN/dE = N_0 (E/E_b)^{-(\Gamma+\beta \log(E/E_b))}$
has been found to be preferable over a power law fit (with $\Delta TS=71.9$).
The spectral fits for the GRBs are performed below 6 GeV and 3 GeV for GRB 090902B and GRB 080916C, respectively. The TS values 
are obtained through a likelihood ratio test comparing a model with background only and a model where a point source was added.}
\label{tab:fit}
\end{center}
\end{table}

\begin{figure}
\epsscale{0.9}
\plotone{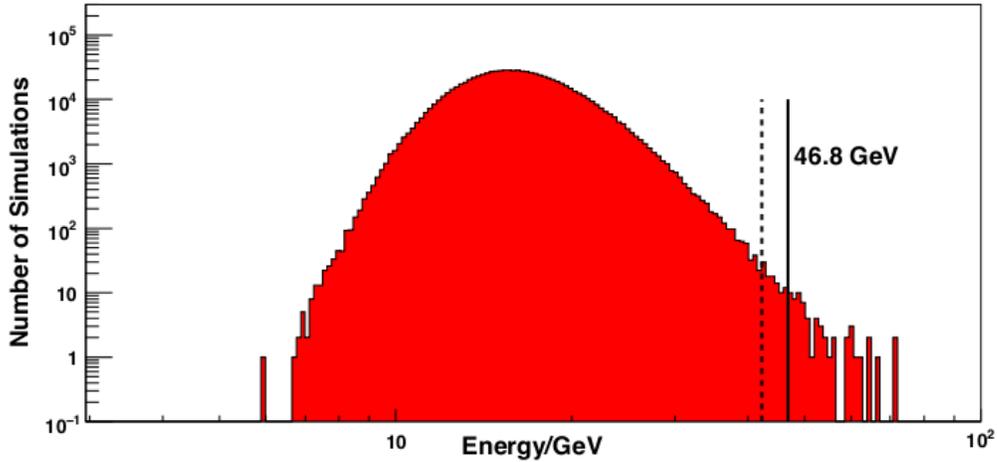}
\caption{Distribution of highest-energy photons obtained from Monte-Carlo
simulations of the source J0808-0751 with the EBL attenuation by
\citet{Stecker06}. $E_{max}$ and $E_{max} - \sigma_{Emax}$ (where  $\sigma_{Emax}$ is the energy uncertainty) are indicated by a solid and dotted vertical black lines, respectively. The probability of detecting a photon with energy equal or greater to  $E_{max} - \sigma_{Emax}$ is equal to
$6.8\times10^{-5}$.}
\label{fig:hep_mcsim}
\end{figure}

\begin{figure}
\epsscale{1.}
\centering
  \begin{minipage}[b]{8cm}
      \includegraphics[width=8.5cm]{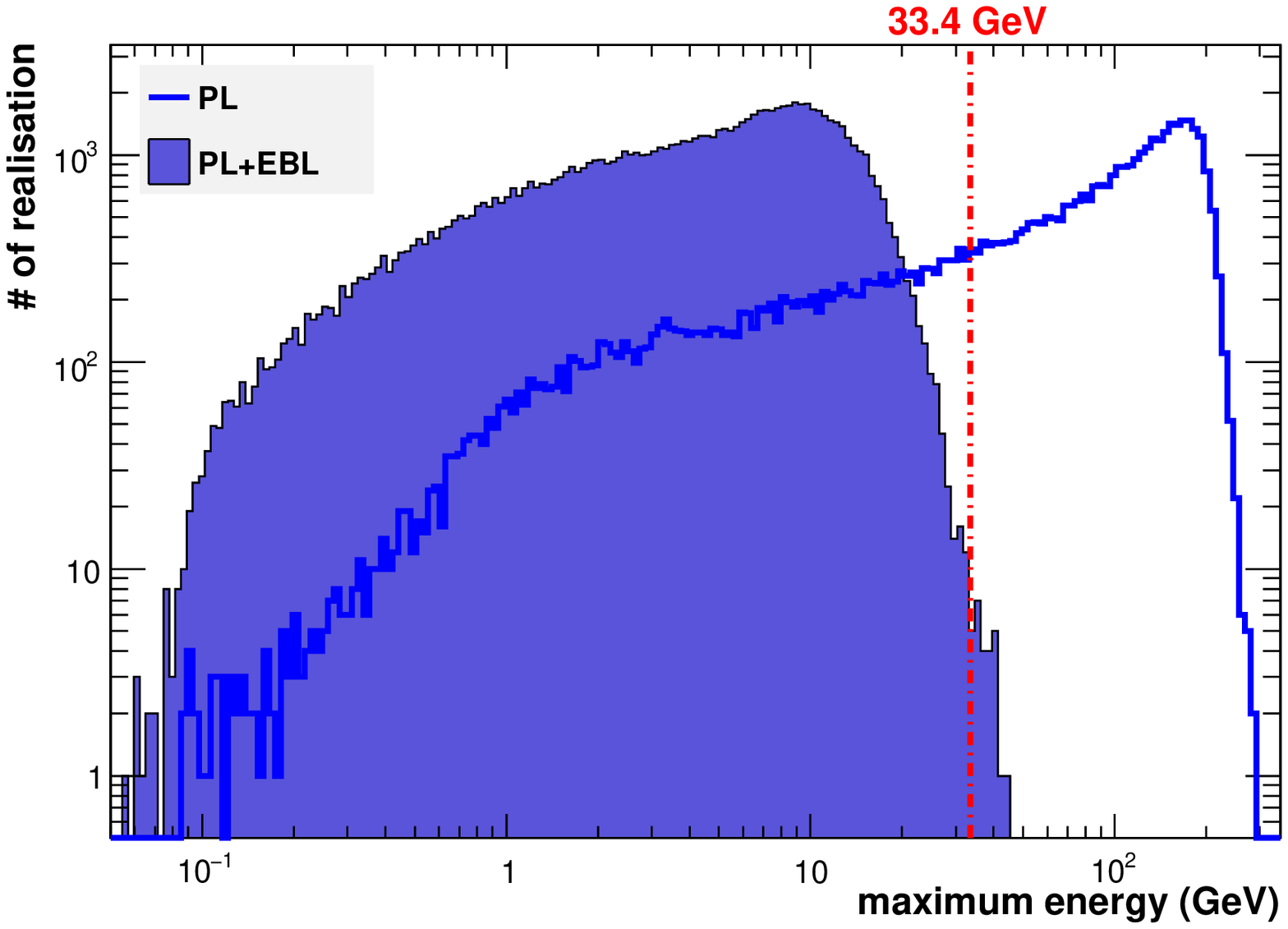}
  \end{minipage}
  \begin{minipage}[b]{7.5cm}
     \includegraphics[width=8.5cm]{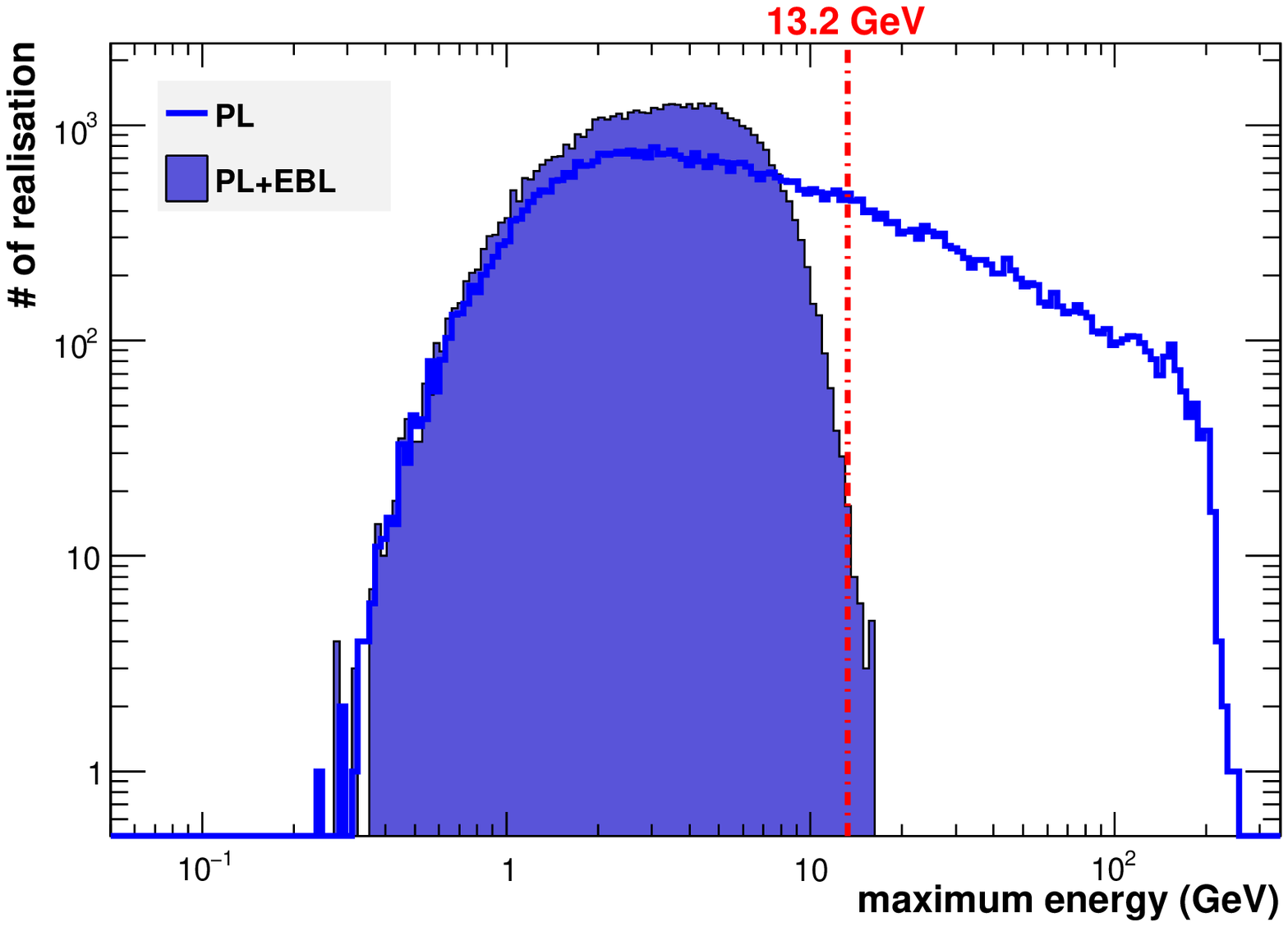} 
  \end{minipage}
\caption{Distributions of the highest energy photons from simulations
performed with estimates of our intrinsic spectra for GRB 080916C
(left panel) and GRB 090902B (right panel), folded with EBL
attenuation calculated using the \citet{Stecker06} baseline model.
The total number of realizations ($10^5$) in both the power-law and power-law convolved with the EBL cases is the same.}
\label{fig:heevt}
\end{figure}

The resulting distribution of the HEP simulated in each case (see
e.g., Figure~\ref{fig:hep_mcsim}) is used to estimate the chance
probability of detecting a photon from the source with energy equal or greater than  $E_{\rm max}$.
We produced 
$\sim 800 000$ and $\sim 100 000$ simulations for each of the HEP sets for
AGN and for GRBs, respectively. Assuming the HEP is indeed from the source, the probability of observing such high energy photon 
given the specific EBL model tested (called $P_{HEP}$) is calculated as the ratio between the number of cases where the HEP energy
 is above $E_{max}$ (actually $E_{max} - \sigma_{Emax}$  given the energy dispersion) and the total number of simulations performed. 
The number of simulations in each case was chosen to reach sufficient statistics at the tail of the distribution where the energy 
of the HEP is measured.
Distributions of the HEP events from these MC simulations for
GRB 080916C and GRB 090902B are shown in Figure~\ref{fig:heevt}.  
The open and filled histograms correspond to the
distributions using the GRB spectra without and with EBL absorption
using the ``baseline model'' of \citet{Stecker06}.

To compute the final probability of rejection for the specific EBL model tested (called $P_{rejection}$), one needs 
to consider the 
fact that the HEP could be a background photon. We compute the probability for this to happen in Table~2 ($P_{bkg}$). 
In the end, 
one can fail to reject the EBL model because the HEP might be a background event or because there is a chance for a 
source photon 
with energy $E_{max}$ not to be absorbed by the EBL so that 
\begin{equation}
P_{rejection} = P_{bkg} + P_{HEP} \times (1-P_{bkg}).
\label{Prej}
\end{equation} 
\noindent In Table~\ref{tab:prob}, we list these 3 probabilities for each of our
most constraining sources. When more than one photon is available for
a given source, the probabilities are combined resulting in
a stronger rejection.  Although $P_{HEP}$ can be quite constraining,
our final significance of rejection is limited by $P_{bkg}$ which is
non-negligible in the case of blazars and which depends on the size of
the region around each source defined {\em a priori} to
look for associated high-energy events (68\% PSF containment radius in
this analysis). A larger  HEP acceptance region (90\% or 95\%
containment radius instead of 68\%) would increase the background
probability $P_{bkg}$ while also adding constraining photons to the
HEP set.  On a source-by-source basis, the rejection probability goes
up or down with increasing radius  depending on the number and
energy of these additional photons, but our overall result remains the same. 
The unbinned likelihood method,
which we describe in the next subsection (\ref{sec:likelihood}),
does not make use of an acceptance radius, and instead makes full use
of available information in the data to systematically calculate
a model rejection probability.

The analysis described in this section was applied to all sources from the HEP set. We find the ``baseline'' 
model of \citet{Stecker06} to be significantly constrained by our observations. Column 5 of
Table~\ref{tab:spectres} shows the optical depth of the ``baseline'' model of \citet{Stecker06} for the HEP
events. Since the
``fast evolution'' model\footnote{The ``baseline'' model considers the case where all galaxy $60\mu$m luminosities
evolved as $(1+z)^{3.1}$ up to $z\leq 1.4$, non-evolving between
$1.4 < z < 6$ and no emission at $z>6$. In contrast, the ``fast evolution'' model assumes a more rapid galaxy luminosity 
evolution: $\propto (1+z)^4$ for $z<0.8$, $\propto (1+z)^2$ for $0.8<z<1.5$, no evolution
for $1.5<z<6$, and no emission at $z>6$. Consequently, for a given redshift the ``fast evolution'' model predicts a higher 
$\gamma$-ray attentuation than the ``baseline'' model.}
of \citet{Stecker06} predicts higher opacities in the LAT energy range at all redshifts, our constraints on this model will naturally be 
higher than the ones found in Table~\ref{tab:spectres} for the ``baseline'' model.

\begin{table}
\begin{center}
\renewcommand{\arraystretch}{1.2}
\begin{tabular}{|c|c|c|c|c|c|cl} 
\hline
Source & z & $E_{max}$ & $\tau (z,E_{max})$ & $\tau (z,E_{max})$ & Number of photons \\ 
       &   & (GeV)     & (F08)       & (St06, baseline)    &  above 15 GeV    \\
\hline\hline
J1147-3812 & 1.05 & 73.7 & 0.40 & 7.1 & 1 \\
J1504+1029 & 1.84 & 48.9 & 0.56 & 12.2  & 7\\
J0808-0751 & 1.84 & 46.8 & 0.52 & 11.7  & 6\\
J1016+0513 & 1.71 & 43.3 & 0.39 &  9.0  & 3\\
J0229-3643 & 2.11 & 31.9 & 0.38 & 10.2 & 1\\
GRB 090902B & 1.82 & 33.4 & 0.28  & 7.7 & 1\\
GRB 080916C & 4.24 & 13.2 & 0.08  & 5.0 & 1\\
\hline
\end{tabular}
\caption{Gamma-ray optical depth to HEP calculated using the EBL model of Franceschini et al (2008; F08)
in comparison to the ``baseline" model of \citet{Stecker06} (St06). Also listed are the number of photons associated to the source which have $\geq 15$ GeV energy and which can potentially constrain EBL models.}
\label{tab:spectres}
\end{center}
\end{table}

\begin{table}
\begin{center}
\begin{scriptsize}
\begin{tabular}{|c|c|c|c|c|c|cl} 
\hline
 &  &  &  & \multicolumn{2}{|c|}{HEP method applied to Stecker 06}  & HEP Rejection \\ 
Source & $z$ & Energy (GeV) & $P_{bkg}$&$P_{HEP}$ & $P_{rejection} $ & Significance\\ 
\hline\hline
J1147-3812 & 1.05 & 73.7&  $7.0\times10^{-4}$ &$1.2\times10^{-4}$ & $8.1\times10^{-4}$  & 3.2 $\sigma$ \\
\hline
J1504+1029 & 1.84 & 48.9 &  $5.6\times10^{-3}$ &$6.7\times10^{-5}$ & $5.7\times10^{-3}$ &  \\
                        &          &  35.1 &  $9.8\times10^{-3}$ & $6.8\times10^{-3}$ & $1.7\times10^{-2}$ & \\
                        &          &  23.2 &  $5.6\times10^{-3}$ & $1.8\times10^{-1}$ & $1.9\times10^{-1}$ & \\
                        &          &           &                                     &  \multicolumn{2}{c|}{Combined $P_{rej}$ = $1.7\times10^{-5}$} & 4.1 $\sigma$                 \\
\hline
J0808-0751 & 1.84 & 46.8 & $1.5\times10^{-3}$ &$1.9\times10^{-4}$ & $1.7\times10^{-3}$&  \\
                        &          &  33.1 &  $2.7\times10^{-3}$ & $3.7\times10^{-3}$ & $6.4\times10^{-3}$ & \\
                        &          &  20.6 &  $6.9\times10^{-3}$ & $2.5\times10^{-1}$ & $2.6\times10^{-1}$ & \\
                        &          &           &                                     &  \multicolumn{2}{c|}{Combined $P_{rej}$ = $2.8\times10^{-6}$} & 4.5 $\sigma$                  \\
\hline
J1016+0513 & 1.71& 43.3& $1.1\times10^{-3}$ &$5.4\times10^{-4}$ & $1.6\times10^{-3}$& \\
                        &          &  16.8 &  $8.2\times10^{-3}$ & $4.9\times10^{-1}$ & $4.9\times10^{-1}$ & \\
                        &          &  16.1 &  $8.2\times10^{-3}$ & $6.5\times10^{-1}$ & $6.5\times10^{-1}$ & \\
                        &          &           &                                     &  \multicolumn{2}{c|}{Combined $P_{rej}$ = $5.3\times10^{-4}$} & 3.3 $\sigma$                 \\
\hline
J0229-3643 & 2.11 & 31.9 & $1.7\times10^{-3}$ &$8.9\times10^{-5}$ & $1.8\times10^{-3}$ & 2.9 $\sigma$\\
\hline
GRB 090902B & 1.82 &33.4 & $2 \times 10^{-6}$ &$2.0\times10^{-4}$ & $2.0\times10^{-4}$  & 3.7 $\sigma$ \\
\hline
GRB 080916C & 4.24 &13.2 & $8 \times 10^{-8}$ & $6.5\times10^{-4}$ & $6.5\times10^{-4}$  & 3.4 $\sigma$  \\
\hline
\end{tabular}
\end{scriptsize}
\caption{Listed are the significance of rejecting the ``baseline'' model (\citet{Stecker06}), calculated using the HEP method as described in Section
\ref{sec:HEP method}.  For completeness, we also report individually the probability of the HEP to be a background event ($P_{bkg}$) and the probability for this HEP not to be absorbed by the EBL if it were emitted by the source ($P_{HEP}$) following Eq.~\ref{Prej}. For those sources with more than one constraining photon, the individual and combined $P_{rejection}$ are calculated.  The ``fast evolution" model by \citet{Stecker06} is more opaque and leads to an
even higher significance of rejection. Applying this method to less opaque models leads to no hints of rejection since the probability $P_{HEP}$ is large in those cases (e.g.~ $\gtrsim 0.1$ for the \citet{Franceschini08} EBL model). Note that a log parabola model was used as the 
intrinsic model for source J1504+1029 since evidence of curvature is observed here 
even below 10 GeV (see Table~\ref{tab:fit}).
}
\label{tab:prob}
\end{center}
\end{table}

\subsubsection{Likelihood method}
\label{sec:likelihood}

This second method to constrain specific EBL models makes use of a
Likelihood-Ratio Test (LRT) technique. This approach compares the likelihood of the
Null-hypothesis model ($L_0$) to best represent the data with the likelihood of a
competitive model ($L_1$).  The test statistic (TS) is defined as
$TS = -2 \times (log(L_{0})-log(L_{1}))$. Following Wilks' theorem
\citep{Wil38}, the TS is asymptotically distributed as $\chi^2_n$
(with $n$ the difference in degrees of freedom between the two models) if the two models
under consideration satisfy the following two conditions \citep{Pro02}:
1) the models must be nested and 2) the null-values of the additional
parameters are not on the boundary of the set of possible parameter
values.

For the LRT we use the power law intrinsic spectrum convolved with the EBL
absorption predicted by the model ($\tau_{mod}$) we are testing,
$\exp[-\alpha\tau_{mod} (E,z)] F_{unabs}(E)$, as the observed flux.
For the Null-hypothesis we set $\alpha = 1$ and we compare it to an
alternative model where $\alpha$ is left as a free parameter, which
therefore has one more degree of freedom than the Null-hypothesis.  In
the absense of any flux attenuation by the EBL, $\alpha = 0$.  Note that
we allow the normalization parameter, $\alpha$ to go to negative
values. This choice, although not physically motivated, allows us to
satisfy the second condition mentioned above. As a consequence the Test-Statistics can simply be converted into a significance 
of rejecting the Null-hypothesis by making use of Wilks' theorem. Because of the lack of information on the intrinsic spectrum 
of a distant source above 10 GeV, we use the (unabsorbed) $\leq 10$ GeV observed spectrum as a reasonable assumption for the
functional shape of the intrinsic source spectrum. A simple power law was found to be a good fit to the $\leq 10$ GeV data for 
the sources listed in Table~\ref{tab:spectres} except in the case of J1504+1029 where a log-parabolic spectrum was preferred. 
We note that if the actual intrinsic curvature is more pronounced than the one found with the best fit below 10 GeV, this would 
only make the results more constraining.

\begin{figure}
\epsscale{0.7}
\plotone{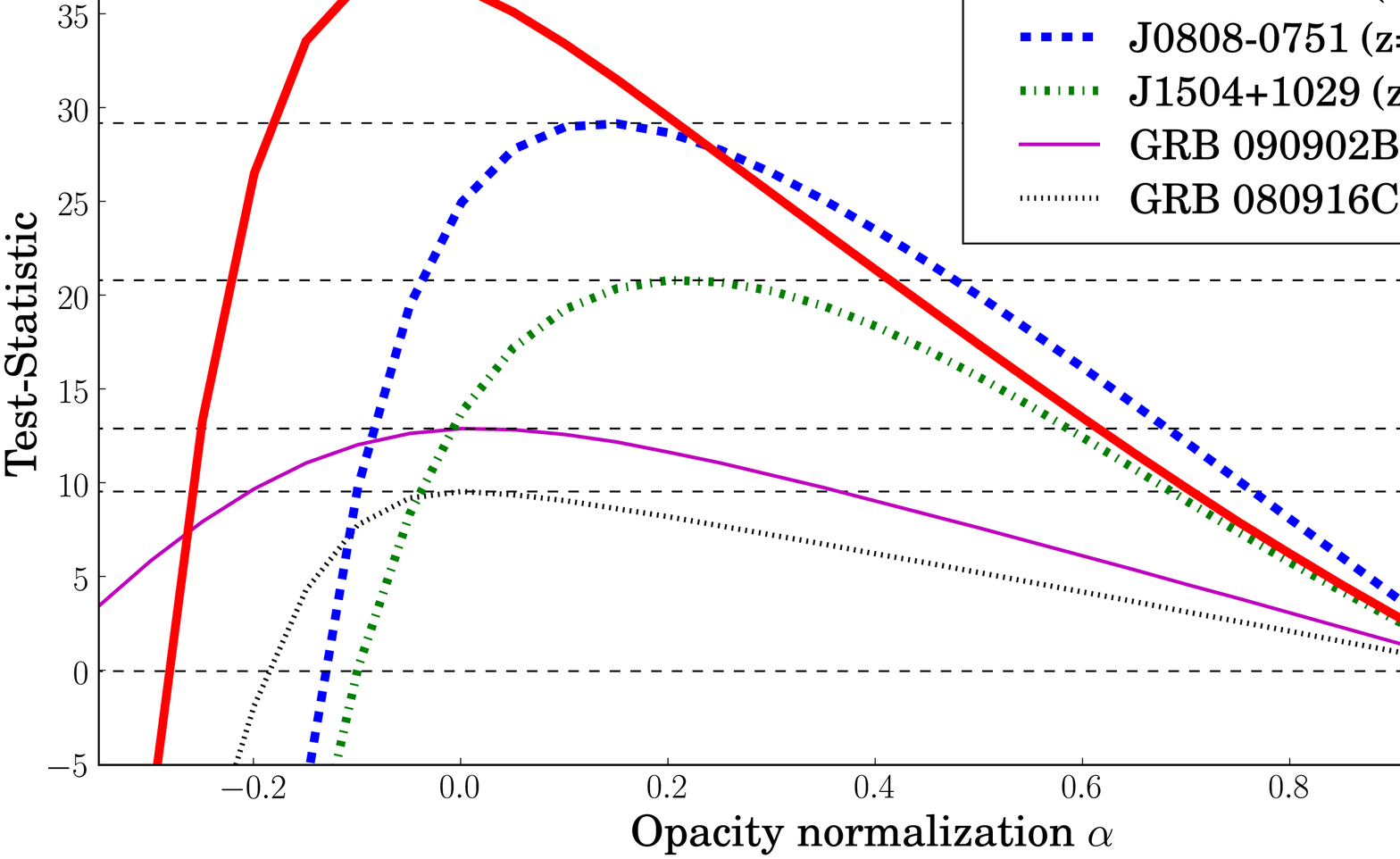}
\caption{Test statistic (TS) as a function of the $\gamma$-ray optical depth normalization
parameter calculated from the likelihood ratio test (LRT) for
J1016+0513, J0808-0751, J1504+1029 and GRBs 090902B and 080916C. The ''baseline'' model of \citet{Stecker06} has been used and the rejection for this model can be directly read out as $\Delta TS$ between $\alpha = 1$ and the best fit $\alpha$ for the source (horizontal dashed line). The confidence interval for the normalization parameter can be obtained using $\Delta TS = CL^2$ where $CL$ is the confidence level.}
\label{fig:TS}
\end{figure}

As we mentioned earlier, although we are considering all EBL models in the literature, we find that our observations are only constraining the most opaque ones.
Figure~\ref{fig:TS} shows the TS value as a function of the optical depth
normalization parameter $\alpha$, for the three most constraining blazars (J1016+0513, J0808-0751, J1504+1029) and the two GRBs (GRBs 090902B and 080916C) when considering the ``baseline'' model of \cite{Stecker06} with the LRT method.  All sources are found to have an optical depth normalization parameter that is consistent with $\alpha \geq 0$ at the $1\sigma$ level which is reassuring as we do not expect a rise in the spectrum on a physical basis. $\sqrt{TS_{max}}$ for $\alpha=1$ corresponds to the rejection significance for the specific model considered.
The most constraining source, J1016+0513,  rejects the Null-hypothesis ($\alpha = 1$, corresponding to the ``baseline'' model of \cite{Stecker06} in this case) with a significance of $\sim 6.0\sigma$.
This source could also constrain the ``high UV model" of Kneiske et al. (2004) with a significance of $3.2\sigma$ although multi-trials effect substantially reduce this significance (see Section~\ref{sec:combined_P}).

\begin{table}{}
\begin{center}
\begin{small}
\begin{tabular}{|c|c|c|c|}
\hline
 & & \multicolumn{2}{|c|}{LRT Rejection Significance} \\
Source & $z$  & pre-trial & post-trial \\
\hline\hline
J1147-3812 & 1.05 & 3.7$\sigma$ & 2.0 $\sigma$ \\
\hline
J1504+1029 & 1.84 & 4.6$\sigma$ & 3.3 $\sigma$ \\
\hline
J0808-0751 & 1.84   & 5.4$\sigma$ & 4.4 $\sigma$ \\
\hline
J1016+0513 & 1.71 & 6.0$\sigma$ & 5.1 $\sigma$ \\
\hline
J0229-3643 & 2.11  &3.2$\sigma$ & 1.2 $\sigma$ \\
\hline
GRB 090902B & 1.82 & 3.6$\sigma$ & 1.9 $\sigma$ \\
\hline
GRB 080916C & 4.24  & 3.1$\sigma$ & 1.0 $\sigma$ \\
\hline
\end{tabular}
\end{small}
\caption{Significance of rejecting the ``baseline'' model (\citet{Stecker06}), calculated using the LRT method described in 
Section \ref{sec:likelihood}.  Again,  the ``fast evolution" model by \citet{Stecker06}  leads to a  high rejection significance 
with two sources (J0808-0751 and J1016+0513) with $> 4 \sigma$ post-trial significance. The post-trial significance is computed 
by taking into account the fact that our analysis is considering $\sim 200$ independent sources.
}
\label{tab:lrt}
\end{center}
\end{table}

As compared to the HEP method, the LRT method incorporates the
possibility of each photon being from the background into the unbinned
maximum likelihood computation.  Thus separate calculations of the
background probablity and corresponding rejection probablity are not
needed.  Also since the LRT method takes into account all high-energy
photons rather than the highest-energy ones in the HEP method, it
gives more constraining results for the EBL model rejection with the
exception of 2 GRBs where the HEP method gives slightly more constraining
results.
Finally, we note that the {\em a priori} choice of the size of the region around each source defined to look for associated high-energy events is a source of systematics for the HEP method ( which uses 68\% PSF containment radius) while it does not affect the LRT method.

\subsubsection{Multi-trial effects and combined probabilities}
\label{sec:combined_P}

Because the search for EBL signatures or rejection of specific EBL
models is performed on all blazars and GRBs detected by the LAT, one
has to consider multi-trials, which is potentially affecting our
analysis.  For independent searches, as is the case here, the
post-trial probability threshold for obtaining a $4\sigma$ result is
$P_{\rm post-trial} = 1-(1- P_{4\sigma})^{1/N_{\rm trials}}$, where
$N_{\rm trials}$ is the number of trials and $P_{4 \sigma}$ is the $4
\sigma$ probability threshold for a single search ($\approx 6.3 \times
10^{-5}$).  In the present case, the LAT AGN catalog that we have
used \citep{1LAC} includes 709 AGNs of which $\sim 200$ have a
sufficiently high redshift ($\sim 100$ with $\gtrsim 10$~GeV photon)
to allow for the testing of EBL attenuation models with their $\g$-ray
spectra.  Only a handful of LAT GRBs were observed with sufficient
statistics to hope to constrain the EBL. In the end, we have
$N_{trials} \sim 200$ which corresponds to a post-trial probability
for a $4 \sigma$ result of $P_{4\sigma,post-trial} \approx 3.17
\times 10^{-7}$. This corresponds to a significance of $\approx 5.11
\sigma$ on an individual source which we will therefore consider as
our threshold for a $4\sigma$ post-trials rejection significance for
any specific EBL model.  This $P_{4\sigma,post-trial}$ threshold was
reached in case of the ~\citet{Stecker06} ``baseline" model for sources
J0808-0751 and J1016+0513 using the LRT method.  Note that J1504+1029 is only
slightly below this threshold.

Combining specific EBL model rejection probabilities from multiple
sources\footnote{since the spectral fits of all the sources we
considered in this analysis are independent to each other.} we get a
much higher rejection significance. For HEP probabilities the
combined rejection significance for the ~\citet{Stecker06} ``baseline"
model is $\approx 8.9\sigma$ ($\approx 7.7\sigma$ without the 2 GRBs)
using Fisher's method in order to combine results from independent  
tests of the same Null-hypothesis \citep{fisher}. For
the LRT method, we add the individual likelihood profiles to derive an
overall profile from which $\sqrt{TS_{max}}$ gives an overall significance of $11.4 \sigma$ for the same EBL model.
Therefore both methods give very large rejection significances even
after taking multi-trial effects into account. Since the~\citet{Stecker06}
``fast-evolution" model gives opacities larger than the ``baseline
model" in the LAT range, both models can be rejected by our analysis
with very high confidence level.  All other models can not be
significantly rejected even after such stacking procedure is applied.

\subsection{Opacity upper limits}

Upper limits on the $\g$-ray optical depth have been evaluated with a method based on the comparison between the 
measured energy spectrum of the source and the unabsorbed spectrum above 10~GeV.
The unabsorbed spectrum, F$_{unabs}$, is assumed to be the extrapolation of the low-energy part, 
 E$<$10~GeV, of the spectrum (F$_{E<10}$), where EBL attenuation is negligible (see Fig.~\ref{fig:tau_vs_energy}), to higher energies.
F$_{E<10}$ is fitted with a power-law or log-parabola function, according to the best TS value.
At high energies, if 
no intrinsic hardening of the spectrum is present, the measured spectrum, F$_{obs}$, at (observed) energy E and the unabsorbed spectrum, F$_{unabs}$, are related by
Eq.~\ref{flux_relation}. The $\gamma$-ray optical depth can therefore be estimated at any given energy as
\be
\tau_{\g\g}(E,z) = \ln[ F_{unabs}(E)/F_{obs}(E) ].
\label{eq:tau}
\ee 
Since F$_{unabs}$ is evaluated assuming no EBL attenuation, it gives a maximum value. 
Therefore the
optical depth, $\tau_{\g\g}(E,z)$ given by Eq.~\ref{eq:tau} could already be considered as an upper limit, 
assuming that the difference between $F_{unabs}(E)$ and $F_{obs}(E)$ is only due to EBL effects.
The fit of both $F_{obs}$ and F$_{E<10}$ are carried out with a maximum likelihood analysis \citep{MattoxLikelihood}
\footnote{{\it gtlike} tool
in the standard {\em Fermi} LAT {\it Science Tools} package provided by the
{\em Fermi} Science Support Center (FSSC)}.

To evaluate $F_{E<10}$ we have assumed a background model including all the point-like sources within $15^\circ$ from 
the source under study and two diffuse components (Galactic and extra-galactic).  
The Galactic diffuse emission is
modeled using a {\it mapcube} function, while a tabulated model is used for
the isotropic component, representing the extragalactic emission as
well as the residual instrumental
background\footnote{http://fermi.gsfc.nasa.gov/ssc/data/access/lat/BackgroundModels.html}.
Both diffuse components are assigned a free normalization for the
likelihood fit.  
In the fit we have considered all the nearby point sources within a 10$^\circ$ radius, 
modeled with a power-law with the
photon index fixed to the value taken from the 1FGL catalog \citep{1FGL} and the
integral flux parameter left free.
The remaining point sources are 
modeled with a power-law with all spectral parameters fixed.  

The source
under study has been fitted with a power-law and a log-parabola with all spectral parameters free.
Among the two, we have chosen the fitted function showing the best TS value. The result is that
for all the sources except J1504+1029 a power-law fit is preferred. 
From the fit results of $F_{E<10}$ we have extrapolated the spectral shape to obtain $F_{unabs}(E)$ above 10~GeV.

A different method has been used to derive the measured flux $F_{obs}$ in selected energy bins.
The whole energy range from 100~MeV to 100~GeV is divided in
equal logarithmically spaced bins requiring in each energy bin a TS value greater than 10: 
2 bins per decade above 10 GeV for J0229-3643,  J1016+0513 and
J1147-3812, 4 bins per decade for J0808-0750 and 5 bins per decade for J1504+1029.
In each energy bin the standard \emph{gtlike} tool has been applied assuming for all the 
point-like sources a simple power law spectrum with photon index fixed to 2.0\footnote{since the energy bin is small enough to assume a flat spectrum.}
The integral fluxes of all point-like sources within $10^{\circ}$
are left as free parameters in the fitting procedure, while the diffuse background components are modeled as described in the previous paragraph.
In this way,
assuming that in each energy bin the spectral shape can be
approximated with a power law, the flux of the source
in all selected energy bins is evaluated.

Once both $F_{unabs}$ and $F_{obs}$ are determined, the maximum $\g$-ray optical depth
in each energy bin can be estimated from Eq.~(\ref{eq:tau}).  
An upper limit on $\tau_{\g\g}(\langle E \rangle,z)$ with 95\% CL in a constraining
energy bin with mean energy $\langle E \rangle$ is then calculated by
propagating the parameter uncertainties in the fitted flux\footnote{It has been veryfied that the statistical errors follow a Gaussian
distribution. The standard error propagation formula has then been
applied.}:

\be
\tau_{\g\g,UL95\%CL}(\langle E \rangle,z) = \ln[ F_{unabs}(\langle E \rangle)/F_{obs}(\langle E \rangle) ] + 2\sigma.
\label{eq:tauUL}
\ee

\noindent We compare these limits with the $\g$-ray optical depths predicted by various EBL models.

\begin{figure}
\epsscale{1.0}
\centering
  \begin{minipage}[b]{7.5cm} 
     \includegraphics[width=6.5cm]{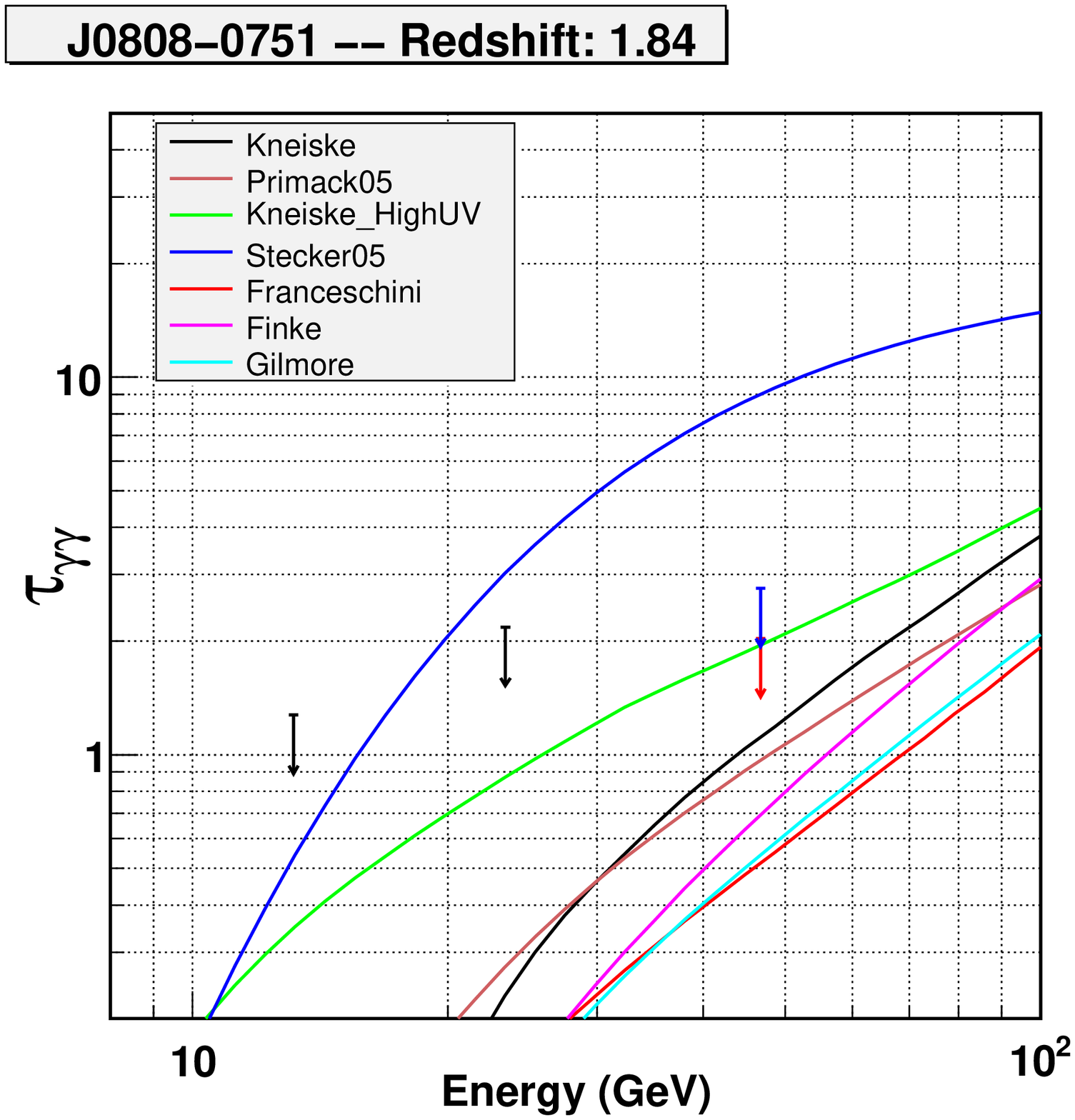}
  \end{minipage}
  \begin{minipage}[b]{7.5cm}
     \includegraphics[width=6.5cm]{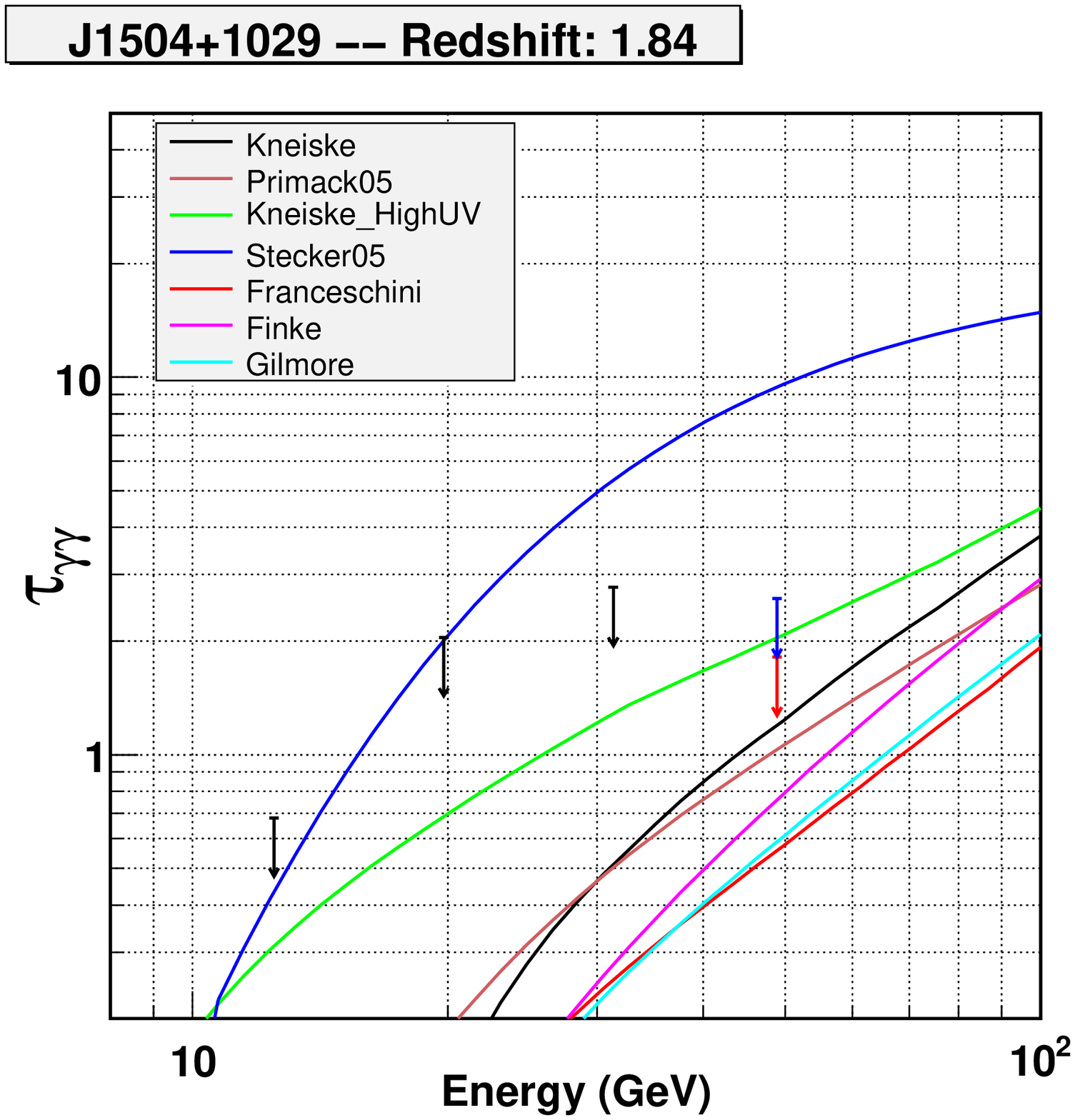} 
  \end{minipage}
  \begin{minipage}[b]{7.5cm}
      \includegraphics[width=6.5cm]{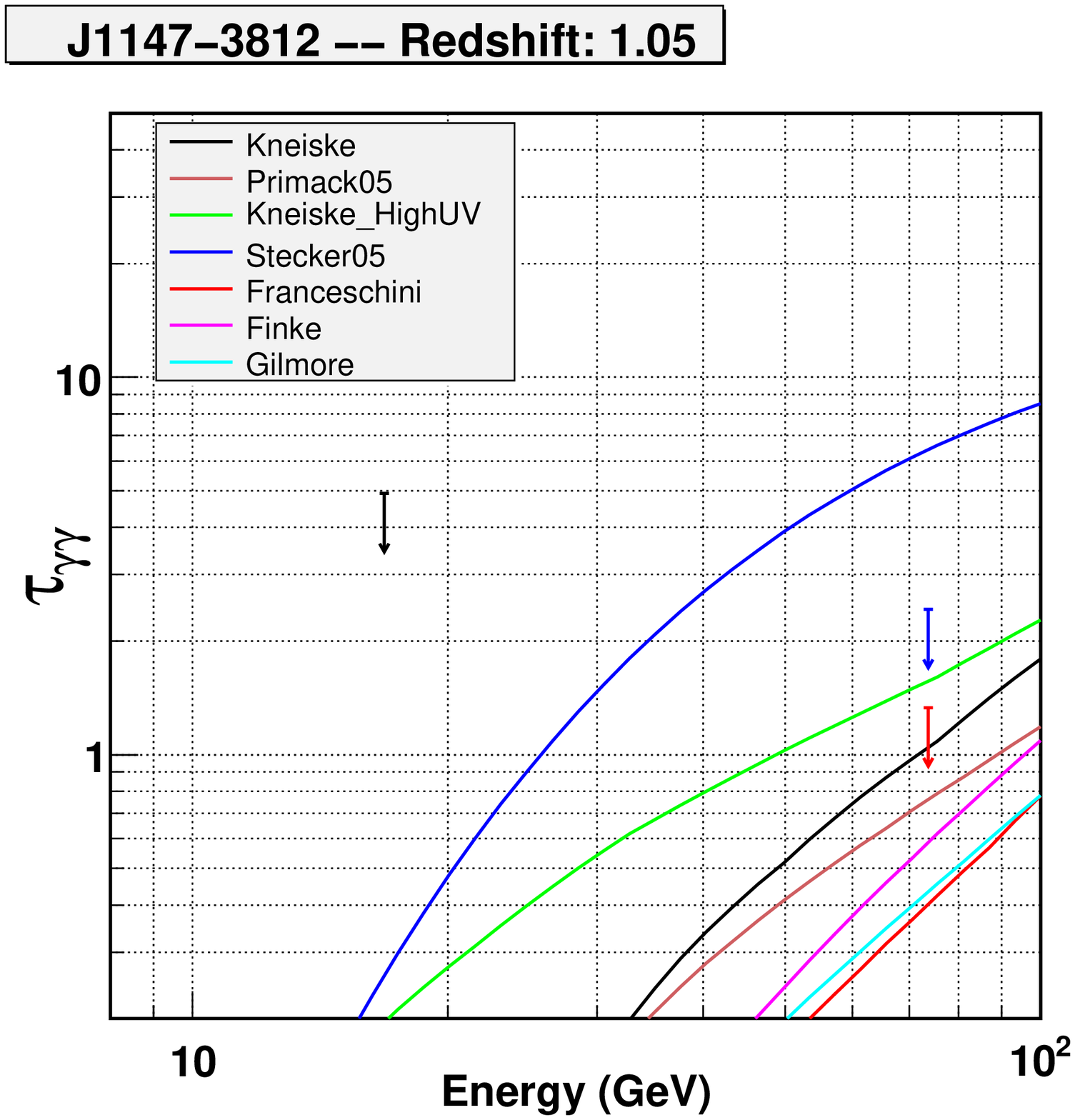}
  \end{minipage}
  \begin{minipage}[b]{7.5cm}
     \includegraphics[width=6.5cm]{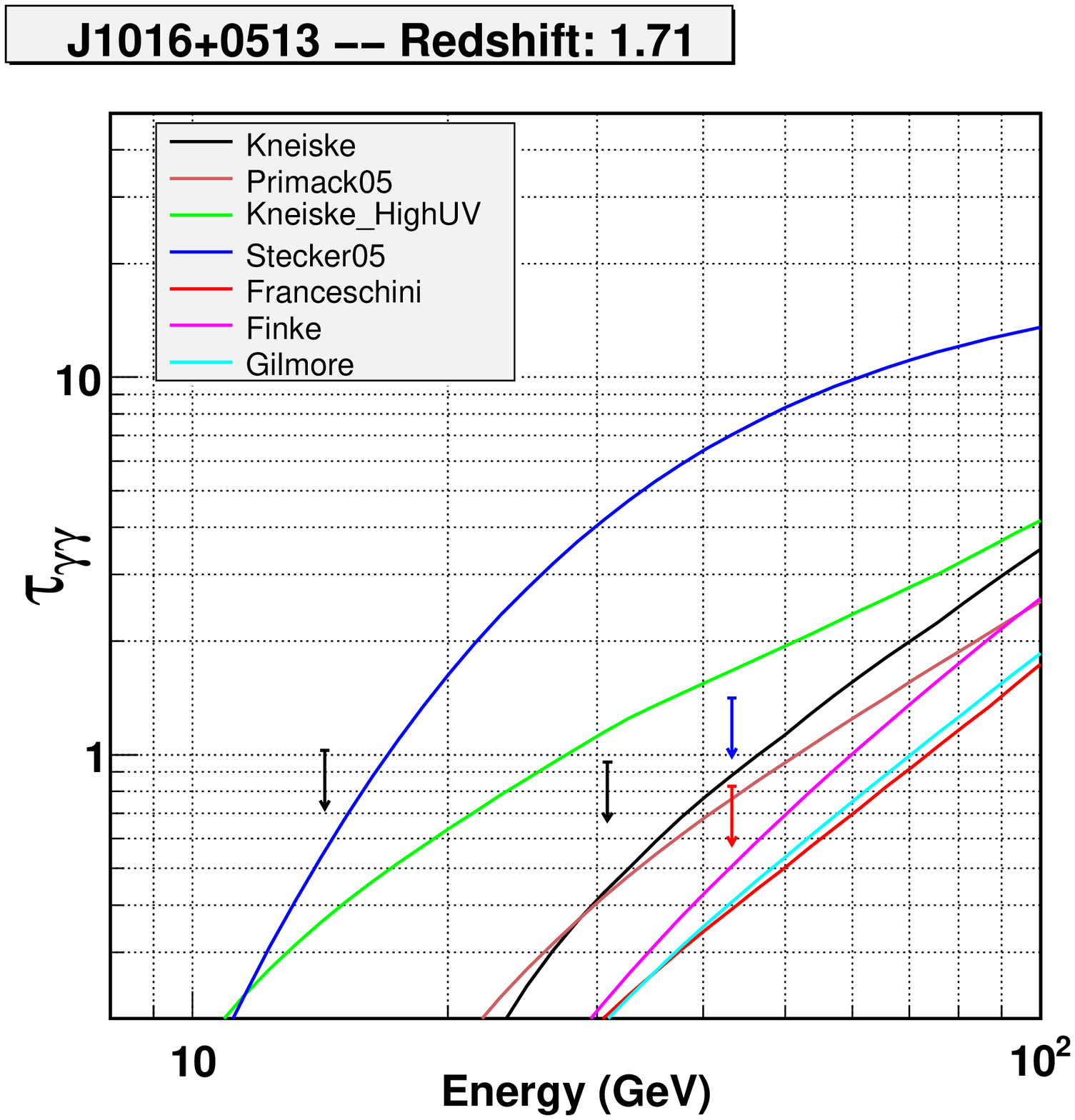} 
  \end{minipage}
\caption{
Derived upper limits for the optical depth of $\g$-rays emitted at z=1.84 (J0808-0751, J1504+1029),
z=1.05 (J1147-3812) and z=1.71 (J1016+0513).
 Black arrows: upper limits at 95\% c.l. in all energy bins used to determine
 the observed flux above 10 GeV. Red arrow: upper
 limits at 95\% c.l. for the highest energy photon. Blue arrow: upper limit
 at 99\% c.l. for the highest energy photon. The upper limits are
 inconsistent with the EBL models that predict the strongest opacity. }
\label{fig:tau_limits}
\end{figure}

In Figure~\ref{fig:tau_limits} we show the upper limits ($95\%$
CL) derived at the mean energy of the bins above 10 GeV for various objects. 
In the highest energy bin the optical depth UL has been evaluated at the highest photon energy
as reported in Table~\ref{tab:spectres}. 
At this energy, the results of the optical depth UL at 99\% CL are also reported (blue arrow).
As an example, consider blazar J0808-0751 at $z=1.84$ shown in the upper left plot: a larger optical depth would require an
intrinsic spectrum that at high energies lies significantly above the
extrapolation obtained from the low energy spectrum.
The figure shows that 
the upper limit rules out those EBL models
that predict strong attenuation. This result is consistent with all
other upper limits derived with this method (see the other plots in Figure~\ref{fig:tau_limits} and summary of the results in Table~\ref{tab:opacityUL}).

\begin{table}
\begin{center}
\renewcommand{\arraystretch}{1.2}
\begin{tabular}{|c|c|c|c|c|cl} 
\hline
Source & z & $E_{max}$ & $\tau_{UL} (z,E_{max})$ & Energy bins \\
 & & & & 10~GeV$-$100~GeV \\
\hline\hline
J1147-3812 & 1.05 & 73.7 &  1.33 & 2 bins/dec \\
J1504+1029 & 1.84 & 48.9 &  1.82 & 5 bins/dec \\
J0808-0751 & 1.84 & 46.8 &  2.03 & 4 bins/dec \\
J1016+0513 & 1.71 & 43.3 &  0.83 & 2 bins/dec \\
J0229-3643 & 2.11 & 31.9 &  0.97 & 2 bins/dec \\
J1012+2439 & 1.81 & 27.6 &  2.41 & 2 bins/dec \\
\hline
\end{tabular}
\caption{Upper limits (95\% c.l.) on the $\g$-ray optical depth for AGN in Table~\ref{tab:spectres}.
The first and second column report the name of the sources and their redshift, 
the third
column the maximum photon energy and the fourth column the optical depth UL
evaluated at 95\% c.l. as $\tau_{UL} = ln[F_{unabs}(E)/F_{obs}(E)] + 2\sigma$., the fourth column the number of energy bins/dec (for E$>$10~GeV) used to evaluate $F_{obs}(E)$.}
\label{tab:opacityUL}
\end{center}
\end{table}

\section{Discussion}
\label{sec:discussion}

Studies with the highest energy extragalactic photons seen by the
{\em Fermi} LAT primarily probe the UV and optical components of the EBL.
The background fields responsible for $\g$-ray attenuation can
evolve strongly with redshift.  In many of the models analyzed in this
paper the EBL intensity can exceed the local value by a factor of 10
or more at redshifts near the peak of star-formation rate density.
The optical depth to $\gamma$ rays from extragalactic sources
 is therefore determined by integrating the EBL intensity
along the line of sight to the source from the observer.

For an interaction angle of $\theta = \pi$, the electron-positron pair
production threshold condition leads to the value of the longest
wavelength photon with which a $\g$-ray emitted at $z_{\rm src}$ with
observed energy $E_{\rm obs}$ can interact (source frame):

\begin{equation}
\lambda_{\rm max}=47.5 \: (1+z_{\rm src}) 
\; \left[{\frac{E_{\rm obs}}{\mbox{GeV}}}\right] \; \mbox{\AA~.}  
\label{max_wavelength}
\end{equation} 

\noindent The equation describes an upper limit on background photon wavelengths
that can contribute to the $\g$-ray optical depth.  Limits for
$\lambda_{\rm max}$ include 
7175 \AA~for blazar J1147-3812, and 4474 \AA~for GRB 090902B and 3286
\AA~for GRB 080916C, based on the highest-energy $\g$-rays seen from
these sources.  In reality, interactions with shorter wavelength
photons are more likely and will contribute more to the optical depth
due to the redshifting of the $\g$-ray during propagation to Earth,
the cross section, which peaks at approximately twice the threshold energy, and
the geometry (interactions at angles of $\sim
90^\circ$ are more likely than head-on interactions as used in
Eq.~\ref{max_wavelength}).

The results of our analysis of the highest energy $\g$-rays from
blazars and GRBs detected by the {\em Fermi} LAT disfavor a UV background intensity
at the level predicted both by the baseline and fast-evolution models of
\citet{Stecker06}, although the LAT observations discussed here do not
constrain the predictions of this work at longer wavelengths.  The two
models of this work are based upon a backwards evolution model of
galaxy formation.  In this scenario, the IR SED of a galaxy is
predicted from its luminosity at 60 $\mu$m.  The locally-determined 60
$\mu$m luminosity function is then assumed to undergo pure luminosity
evolution following a power law in $(1+z)$.  Optical and UV
luminosities, relevant to {\em Fermi}'s extragalactic observations, are then
determined by analytic approximation to the SEDs in \citet{Salamon98}
and are normalized to the short wavelength portions of the IR SEDs.
These models do not include absorption of UV light by dust in
star-forming regions and the interstellar medium of galaxies, which
may partially account for the high background in this model.  While
this model does account for redshift evolution in the UV-optical SEDs
of galaxies, it does not allow for any evolution in the IR emission to
which these SEDs are normalized.  As mentioned by \citet{Stecker06},
this is another factor which could result in overpredicted UV
emission.

Emissivity at UV wavelengths is closely tied to the global
star-formation rate density.  Because the models of \citet{Stecker06}
are not derived from an assumed function for the star-formation rate,
limits on the UV emissivity in this case cannot be used to directly
constrain star-formation.  We do not find that our results
are conclusively in disagreement with the `best-fit' model of \citet{Kneiske04}. 
In these models, the
optical-UV EBL is based upon a Salpeter IMF and a star-formation rate
density that peaks at $z\sim$1.25, with a value of $\sim 0.2$
M$_\odot$ Mpc$^{-3}$ yr$^{-1}$, and falls slowly towards higher
redshift.  In the high-UV model, ultraviolet flux is boosted by a
factor of 4 above the level of the best-fit model, greatly enhancing
the opacity for $\g$-rays at energies below about 200 GeV.  A star
formation history of the magnitude required to produce the background
in the high-UV model would be above essentially all estimates of the
global star formation rate (see for example
\citet{Hopkins&Beacom06}). All other EBL models are of such low density in the
UV range that they can not be constrained by the data presented in this work.
Although the results
of our analysis can not yet place any stringent upper limits on the cosmological
star-formation history that are competitive with current observational
estimates, future prospects for probing low density UV models of the EBL 
by means of improved methods and enlarged GeV photon data sets
may be promising.

High-energy $\g$ rays that are absorbed by the EBL photons can
initiate a pair cascade by subsequent Compton scattering of the CMB
photons by the pairs.  In case the intergalactic magnetic field (IGMF)
is very weak, so that the pairs do not deflect out of our line of
sight, this cascade radiation component can be
detectable~\citep{Plaga95}.  Calculations of such cascade signatures have been
carried out for AGNs~\citep[see e.g.][]{Dai02,Murase08,Essey10} and
for GRBs~\citep[see e.g.][]{DL02,Razzaque04,Takahashi08} and found to
compensate for a large portion of the flux that is absorbed in the EBL.
If blazars or GRBs are sources of ultrahigh energy cosmic rays (UHECRs; \citet{Waxman}), then photohadronic 
interactions by protons during their propagation in the background light can also 
induce a high-energy cascade signature that would 
form appreciable high-energy emission, provided the IGM magnetic field is sufficiently small
\citep{Essey10}.
However recent flux upper-limits calculated in the {\em
Fermi} LAT range from TeV blazars 1ES 0347-121 and 1ES 0229+200
constrain the IGMF to be $> 3\times 10^{-16}$~G~\citep{Neronov10}.
Such a strong field reduces the cascade flux significantly (because the emission
becomes essentially isotropic due to the large deflection angles) and the
contribution to the observed flux is likely to be small.  
Furthermore, since the constraining blazar sample consists of FSRQs only, which seem weak TeV emitters,
any of their reprocessed emission can only be small also.

Exotic scenarios involving oscillation between $\g$-rays and axionlike
particles, while propagating in the Galactic magnetic field, from
distant sources may produce observable signatures in the TeV
range~\citep[see e.g.][]{axions1,axions2}.  However the effect may not
set in for typically assumed IGMF values or likely to be too small to
make up for EBL flux attenuation in the $\lesssim 100$~GeV
range~\citep[see e.g.][]{axions3}

\section{Conclusion}
\label{sec:conclusion}

Using the high-energy 11-month photon data set collected by {\em Fermi}
from distant blazars, and two GRBs we have (i) placed upper limits on the opacity of
the Universe to $\g$ rays in the $\sim$10--100~GeV range coming
from various redshifts up to $z\approx 4.3$; and (ii) ruled out an EBL
intensity in the redshift range $\sim$ 1 to 4.3 as great as that predicted by
\citet{Stecker06} in the ultraviolet range  
at more than $4\sigma$ post trials in two independent sources (blazars). The overall rejection significance is found to be $>10 \sigma$ post trials therefore making this result very robust. Our most 
constraining sources are blazars
J1504+1029, J0808-0751 and J1016+0513 with $(z, \langle E_{max} \rangle)$
combinations of (1.84, 48.9~GeV), (1.84, 46.8 GeV) and (1.71, 43.3 GeV), respectively.
Although a likelihood ratio analysis of the latter source indicates
that the sensitivity of our analysis method is approaching the EBL flux level of
the ``high UV model'' of Kneiske et al (2004), multi-trial effects markedly reduced
the rejection significance.
The two most constraining GRBs are GRB 090902B and GRB 080916C, both
of which rule out the ``baseline'' EBL model of \citet{Stecker06} in the UV energy range at more than
3$\sigma$ level. The ``fast evolution'' model of \citet{Stecker06} predicts higher opacities in the LAT energy range at
all redshifts, and therefore is also ruled out. Together with the results from VHE observations (e.g., \citet{aharonian07,Mazin07})
the models by \citet{Stecker06} seem now disfavored in the UV and mid-IR energy range.
We have also calculated model-independent optical depth
upper-limits $\tau_{\g\g, \rm UL} (z, \langle E_{max} \rangle)$
at 95\% CL in the redshift $z\simeq 1-2.1$ and $E_{\rm max}\approx 28-74$GeV ranges.

As the high-energy photon data set collected by {\em Fermi} grows in the future
and more blazars and GRBs are detected at constraining energies, the
$(E, z)$ phase space that constrains $\tau_{\g\g}$ will become more
populated.  
This will provide us with unique opportunities to
constrain the opacity of the Universe to $\g$-rays over a large energy
and redshift range, and eventually help us further understand the evolution of the
intensity of the extragalactic background light over cosmic time.


\acknowledgments
The \textit{Fermi} LAT Collaboration acknowledges generous ongoing support
from a number of agencies and institutes that have supported both the
development and the operation of the LAT as well as scientific data analysis.
These include the National Aeronautics and Space Administration and the
Department of Energy in the United States, the Commissariat \`a l'Energie Atomique
and the Centre National de la Recherche Scientifique / Institut National de Physique
Nucl\'eaire et de Physique des Particules in France, the Agenzia Spaziale Italiana
and the Istituto Nazionale di Fisica Nucleare in Italy, the Ministry of Education,
Culture, Sports, Science and Technology (MEXT), High Energy Accelerator Research
Organization (KEK) and Japan Aerospace Exploration Agency (JAXA) in Japan, and
the K.~A.~Wallenberg Foundation, the Swedish Research Council and the
Swedish National Space Board in Sweden. Additional support for science analysis during the operations phase is gratefully
acknowledged from the Istituto Nazionale di Astrofisica in Italy and the Centre National d'\'Etudes Spatiales in France. The \textit{Fermi} GBM collaboration acknowledges 
support for GBM development, operations and data analysis from NASA in the US and BMWi/DLR in Germany. L. Reyes acknowledges support by the
 Kavli Institute for Cosmological Physics at the University of Chicago through grants NSF PHY-0114422 and NSF PHY-0551142 and an 
 endowment from the Kavli Foundation and its founder Fred Kavli. AR acknowledges support by Marie Curie IRG grant 248037 within the FP7 Program. Furthermore, helpful comments from the referee are acknowledged.

\end{document}